\documentclass[sigconf, nonacm]{acmart}
\usepackage{pifont}
\usepackage{comment}
\usepackage{float}
\usepackage[utf8]{inputenc}
\usepackage{balance}
\usepackage{algpseudocode}
\usepackage{subcaption}
\usepackage{tabularx}
\usepackage[export]{adjustbox}
\usepackage{diagbox}
\usepackage{hyperref}
\usepackage[noabbrev, capitalise, nameinlink]{cleveref}
\usepackage{xcolor,colortbl}
\usepackage[vlined,noend,linesnumbered,ruled,resetcount]{algorithm2e}
\let\oldnl\nl
\newcommand{\nonl}{\renewcommand{\nl}{\let\nl\oldnl}}

\usepackage{amsmath,amsfonts,amsthm,bm}
\usepackage{thmtools}
\usepackage{thm-restate}

\declaretheorem[]{theorem}
\declaretheorem[]{problem}
\declaretheorem[]{lemma}

\declaretheorem[]{corollary}
\declaretheorem[]{definition}

\def\eqref#1{equation~\ref{#1}}

\def\1{\bm{1}}

\def\vq{{\bm{q}}}

\def\vx{{\bm{x}}}

\def\mA{{\bm{A}}}
\def\mB{{\bm{B}}}
\def\mC{{\bm{C}}}
\def\mD{{\bm{D}}}

\def\mW{{\bm{W}}}

\DeclareMathAlphabet{\mathsfit}{\encodingdefault}{\sfdefault}{m}{sl}
\SetMathAlphabet{\mathsfit}{bold}{\encodingdefault}{\sfdefault}{bx}{n}

\def\gC{{\mathcal{C}}}

\def\gM{{\mathcal{M}}}

\def\gP{{\mathcal{P}}}
\def\gQ{{\mathcal{Q}}}

\def\gS{{\mathcal{S}}}

\DeclareMathOperator{\Error}{Error}
\DeclareMathOperator{\Lap}{Lap}

\newcommand{\new}[1]{{#1}}

\newcommand\vldbpagestyle{empty} 

\begin{document}
\title{Multi-Analyst Differential Privacy for Online Query Answering}

\author{David Pujol}
\affiliation{%
  \institution{Duke University}
}
\email{dpujol@cs.duke.edu}

\author{Albert Sun}
\affiliation{%
  \institution{Duke University}
}
\email{albert.sun310@duke.edu}

\author{Brandon Fain}
\affiliation{%
  \institution{Duke University}
}
\email{btfain@cs.duke.edu}

\author{Ashwin Machanavajjhala}
\affiliation{%
  \institution{Duke University}
}
\email{ashwin@cs.duke.edu   }

\begin{abstract}
\label{abstract}
Most differentially private mechanisms are designed for the use of a single analyst. In reality, however, there are often multiple stakeholders with different and possibly conflicting priorities that must share the same privacy loss budget. This motivates the problem of equitable budget-sharing for multi-analyst differential privacy.  Our previous work defined desiderata that any mechanism in this space should satisfy and  introduced methods for  budget-sharing in the offline case where queries are known in advance.  \par 
We extend our previous work on multi-analyst differentially private query answering to the case of online query answering, where queries come in one at a time and must be answered without knowledge of the following queries. We demonstrate that the unknown ordering of queries in the online case results in a fundamental limit in the number of queries that can be answered while satisfying the desiderata. In response, we develop two mechanisms, one which satisfies the desiderata in all cases but is subject to the fundamental limitations, and another that randomizes the input order ensuring that existing online query answering mechanisms can satisfy the desiderata.

\end{abstract}

\maketitle

\pagestyle{\vldbpagestyle}




\section{Introduction}

\label{Introduction}

Analysis of sensitive information about individuals is essential for many research tasks. However since such analyses are often made public this may come at the cost of individual privacy \cite{haney2017,machanavajjhala_privacy_on_the_map,vaidya_2013}. Differential Privacy (DP) \cite{dwork2014:textbook,DiffPriv} is often considered the gold standard of privacy protection. Differentially Private mechanisms often add randomized noise to data to hide the presence of an individual record while still preserving aggregate information. However, due to the fundamental law of information recovery \cite{dinur_nissim_2003}, answering an unbounded amount of queries, even under Differential Privacy will eventually allow an attacker to accurately reconstruct the underlying dataset. Because of this, data curators must bound the amount of information released. In differential privacy this is captured by setting a bound on the privacy loss parameter, $\epsilon$, often called the privacy loss budget or just simply the privacy budget. This turns any query answering problem into a resource allocation problem. Given a fixed $\epsilon$ how much of the privacy budget should be spent on each query and how will this budget be allocated across the interests of multiple stakeholders?\par 

We study the Multi-Analyst Differential Privacy problem introduced in \cite{Waterfilling}. In this setting, there are multiple stakeholders all interested in a single dataset. The data curator must answer all of the analysts' queries while limiting the total privacy budget across all analysts to be bounded by $\epsilon$. 
Each analyst is entitled to a specific share of the privacy budget.
Since the privacy budget is finite and must be spent to answer queries, we treat it as a fundamental system resource to be distributed across the analysts. \par

 In order to ensure acceptable utility for all analysts, \cite{Waterfilling} considers several additional desiderata that any mechanism in this space should satisfy. They introduce the hypothetical collective where an analyst has the choice of either answering their queries independently with their own share of the privacy budget or joining a group of other analysts and sharing their privacy budgets under a joint mechanism. 
 First, the mechanism must satisfy the \textbf{Sharing Incentive}, meaning that the joint mechanism must offer each analyst at least as much utility as they would receive if they took their share of the privacy budget and answered their queries independently. This protects analysts joining the mechanism ensuring that their utility can only increase by so doing.  A mechanism should also satisfy \textbf{Non-Interference} or the stronger property of \textbf{Analyst Monotonicity}, both stating (at a high level) that the addition of a new analyst to the collective (along with their respective resources) must not cause any analyst already in the collective to suffer a loss in utility. This protects analysts already participating in the joint mechanism, ensuring that the addition of new participants can only increase their utility.

 \par 

Previous work \cite{Waterfilling} considers the offline case where all queries are presented ahead of time and a mechanism outputs all query answers at once. This offline query answering model is equivalent to one time data release such as the data products released by the US Census Bureau \cite{CensusTopDown}. In this paper, we consider the extension to the online query answering model, where queries are asked one at a time without prior knowledge of the entire query sequence.

\subsection{Contributions}
We consider the extension to the Online setting where queries are asked one at a time and the number of queries being asked is not known by the mechanism. This allows for cases where the workload is not known in advance and queries may be sporadic. 
 Our contributions are as follows.
\begin{itemize}
    \item In \cref{Sec:problem}  we extend the multi-analyst differentially private query answering problem to the online setting where queries are in the form of an ordered sequence and must be answered one at a time. 
    \item We show experimentally in \cref{sec:motivation} that existing online mechanisms fail to satisfy the sharing incentive or non-interference and the disparities between analysts grows as the distribution of queries becomes skewed towards one analyst.
    \item \new {We prove \cref{Thrm:SharingIncentiveUpper} which states that there is a fundamental limit to the number of queries a mechanism can answer while still satisfying the sharing incentive and that this limit scales polynomially in the number of analysts.}

\item On the positive side we provide two solutions to the online multi-analyst differential privacy problem. First we introduce a new mechanism in \cref{Sec:CacheAndReconstruct} called \textbf{Seeded Cache and Reconstruct}, an online multi-analyst query answering algorithm that satisfies differential privacy, sharing incentive, analyst monotonicity, and non-interference. This mechanism is subject to the upper bound on queries implied by  \cref{Thrm:SharingIncentiveUpper} but requires no modification to the online query answering model.

\item In \cref{Sec:RandomOrdering} we introduce the \textbf{Query Scheduler}, a method of circumventing the limitations of \cref{Thrm:SharingIncentiveUpper} while still satisfying the sharing incentive by enforcing a structure on the order in which analysts ask queries. We show, by a reduction to the well known Counting Coupons problem, that by randomly selecting an analyst to answer one of their queries at each time step any efficient-enough algorithm satisfies the sharing incentive.

\item Finally we show that this result generalizes to cases where the analysts are not chosen exactly uniformly at random and the privacy budget shares are not exactly equal. 
\end{itemize}

\section{Background}
\label{background}
\noindent\textbf{Data Representation}
Like previous work \cite{HDMM,Matrix,Matrix10,PMW},
we consider databases where each individual corresponds to exactly one tuple. The algorithms considered use a vector representation of the database denoted $\vx$. More specifically, given a set of predicates $ \mathcal{B} = \{\phi_1 \dots \phi_k\}$, the original database $D$ is transformed into a vector of fractional counts $\vx^D $ where $\vx_i^D$ is the fraction of records in $D$ which satisfy $\phi_i$. We denote the length of the data vector as $|x|$ (often referred to as the dimension of the database), the number of individuals in the database $n$, and we will use the notation $\vx$ to refer to the vector form of database $D$. 
\par
\noindent\textbf{Predicate counting queries}
 are a versatile and powerful class of queries that count the number of tuples satisfying a logical predicate. 
 A predicate corresponds to a condition in the \textbf{WHERE} clause of an SQL query. A predicate counting query is one of the form \textbf{SELECT Count (*) FROM R WHERE $\phi$}. Workloads of counting queries can express a rich and powerful set of queries such as histograms, range queries, marginals, and datacubes. Like databases, a predicate counting query can be represented as a $|x|$-length vector $\vq$ such that the answer to the query is $\vq^T\vx$.

\noindent\textbf{Differential Privacy}
 \cite{dwork2014:textbook,DiffPriv} is a formal model of privacy that guarantees each individual that any query computed from sensitive data would have been almost as likely as if the individual had opted out. More formally, Differential Privacy is a property of a randomized algorithm which bounds the ratio of output probabilities induced by changes in a single record. 
\begin{definition}[Differential Privacy]
A rand\-om\-ized mechanism $\gM$ is $(\epsilon$,$\delta$ )-differentially private if for two neighboring databases $D$, and $D'$ which differ in at most one row, and any outputs $O \subseteq Range(\gM)$: the following holds:
\begin{equation*}
\Pr[ \gM(D) \in O] \leq \exp(\epsilon) \times \Pr[ \gM(D') \in O] + \delta
\end{equation*}
\end{definition}
 
The parameter $\epsilon$ often called the privacy budget quantifies the privacy loss. $\delta$ can be seen as the probability of catastrophic failure, for this reason, $\delta$ is usually set to be negligibly low.  \par The Laplace Mechanism is a differentially private primitive which is utilized in many differentially private mechanisms. 
\begin{definition}[Laplace Mechanism]
Given a query vector $q$, the randomized algorithm which outputs the following vector is $\epsilon$-differentially private \cite{dwork2014:textbook}. 
\begin{equation*}
q\vx + \Lap\left(\frac{\|q\|_1}{\epsilon}\right)
\end{equation*}
\end{definition}
Where $\|q\|_1$ is the maximum L1 column norm of $q$, often called the sensitivity of a query, and $\Lap(\sigma)$ denotes a sample from a Laplace distribution with mean $0$ and scale $\sigma$.  

While the sensitivity of a query determines how much noise is necessary, for simplicity of analysis we will consider in this work only queries of sensitivity 1. Any linear counting query can be reduced to a sensitivity of 1 by multiplying it with a  normalizing constant.  
Differentially private releases compose with each other in that if there are two private releases of the same data with two different privacy budgets the amount of privacy lost is equivalent to the sum of their privacy budgets. More formally we have the following. 
\begin{theorem}[DP composition \cite{dwork2014:textbook}]
\label{thrm:composoition}
Let $\gM_1$ be an $\epsilon _1$-differentially private algorithm and $\gM_2$ be an $\epsilon _2$-differentially private algorithm. Then their combination defined to be $\gM_{1,2}(x) = (\gM_1(x), \gM_2(x))$ is $\epsilon _1 + \epsilon _2$-differentially private
\end{theorem}
This allows the privacy budget to be split across multiple mechanisms and have the overall privacy loss quantified across all the mechanisms. In this way, the privacy budget acts like a finite resource that can be spent on various individual tasks.
\noindent\textbf{Private Multiplicative Weights} \cite{PMW} (PMW) is an online differentially private mechanism that creates a synthetic database as queries are being answered. At each time step if the synthetic database can be used to answer an individual query with high enough accuracy then the synthetic database is queried, resulting in no privacy loss. If the synthetic database cannot be queried accurately then the true database is queried using the Laplace Mechanism and the noisy answer is then used to update the synthetic database. PMW is $( \epsilon, \delta)$-differentially private and satisfies $(\alpha, \beta, k)$
- accuracy, meaning that it can answer $k$ queries with error under
$\alpha = O\left(\frac{((\log(d) + \log(1/\beta)) \log^{1/4}(|D|)  \log(1/\delta)}{\epsilon \sqrt{n}} \right)$ with probability $1-\beta$.   \par 

\noindent\textbf{The Matrix Mechanism} \cite{Matrix,HDMM} is an offline query answering mechanism for answering workloads of queries denoted $\mW$. The mechanism creates an alternative "strategy workload" denoted $\mA$ to answer directly and reconstructs the queries in $\mW$ from $\mA$. The error of the matrix mechanism is as follows. 
\begin{equation} \label{eqn:error}
\Error(\mW,\mA, \epsilon) = \frac{2}{\epsilon ^2} \| \mA\|_1^2 \| \mW\mA^+ \|^2_F    
\end{equation}
This will be necessary to evaluate the error of mechanisms later. We will also rely on results initially from \cite{Waterfilling} which are as follows. 
\begin{lemma} \label{lemma:diagonal}
For any workload matrix $\mW$ and any strategy $\mA$ \begin{equation*}
    \left\| \mW(\mD\mA)^+\right\|_F \leq \left\| \mW\mA^+\right\|_F
\end{equation*} where $\mD$ is a diagonal matrix with all diagonal entries greater than or equal to 1 and $\mA$ is a full rank matrix.
\end{lemma}

\begin{lemma} \label{lemma:rows}
Let $\tilde{\mA}$ be the original strategy matrix $\mA$ with additional queries (rows) added to it. We can write this as a block matrix as  $ \tilde{\mA}  = \begin{bmatrix}
\mA \\
\mC
\end{bmatrix} $ 
Where $\mC$ are the additional queries.
For any workload $\mW$ and any strategy $\mA$
\begin{equation*}
    \left\| \mW\tilde{\mA}^+ \right\|_F \leq \left\| \mW\mA^+ \right\|_F
\end{equation*} 
\end{lemma}

Pujol et al. \cite{Waterfilling} showed that when using the matrix mechanism reconstruction step, for any strategy matrix $\mA$ the addition of more queries with additional privacy budget (\cref{lemma:rows}) or additional privacy budget for existing queries (\cref{lemma:diagonal}) will always result in a reconstruction with lower expected error. These are results that we rely on heavily in the construction of the Seeded Cache and Reconstruct mechanism in \cref{Sec:CacheAndReconstruct}.

\noindent\textbf{Counting Coupons}
For results in \cref{Sec:RandomOrdering} we rely on a reduction to the classic combinatorial problem of Counting Coupons.

The traditional Counting Coupons problem has an urn with $k$ unique coupon types from which we sample with replacement. The goal is to calculate $T_k$, the number of coupons that need to be sampled in expectation in order to have at least one of each coupon type. We can generalize this problem in two ways. First, suppose we have a vector $\vec{m}$ indicating we need at least $m_i$ copies of type $i$. Second, suppose we have a vector $\vec{x}$ indicating that there are $x_i$ coupons of type $i$ in the urn. The more general problem is to find $T_k(\vec{m},\vec{x})$, the expected number of samples with replacement needed to obtain $m_i$ copies of each type $i$, given $x_i$ of that type in the urn. We use a scalar for $\vec{x}$ or $\vec{m}$ to denote a length $k$ constant vector of that value repeated. \par 
We will rely on several results from the literature namely the following theorems. 
\begin{theorem}[Coupon Collecting with Non-Uniform Probability \cite{shank_yang_2013}]
\label{Coupons:General}
Let $k$ be the number of unique coupons, $m$ be the vector representing the quotas for each coupon and $p$ be the vector representing the number of each coupon in the urn.
Let $P_k = \sum_{i=0}^k \vec{p}_i$ and let $S_{m_i}(t) = \sum_{k=0}^{m_i-1} \frac{t^k}{k!}$, the first $m_i$ terms of the Taylor series expansion of $e^t$. Then the expected number of samples required to collect $m_i$ coupons of each type denoted $T_n(\vec{m},\vec{p})$ is as follows
\begin{equation}
    T_k(\vec{m},\vec{p}) =  X_n \int_0^{\infty} \left(1-\prod^n_{i=1}(1-S_{\vec{m}_i}(p_it))e^{-\vec{p}_it}) \right) dt
\end{equation}

\end{theorem}
Likewise, we have the special case where all coupons are equally likely and $\vec{m}$ is a constant.
\begin{corollary}[Coupon Collecting with Uniform Probability \cite{DoubleDixie}]
\label{Coupons:Uniform}
\begin{align}
    T_k(m,1) = k \int_0^{\infty} \left(1- \prod_{i=1}^k ( 1- S_{m_i}(t)e^{-t})\right) dt =  \\
k( \log(k) +  (m - 1)\log(\log(k)) + o(1)) 
\end{align}

\end{corollary}

Using \cref{Coupons:General} and \cref{Coupons:Uniform} we can then construct the following theorem from \cite{shank_yang_2013}. 
\begin{theorem}[Coupon Collecting Upper Bound \cite{shank_yang_2013}]
\label{Coupons:Upper}
Let $p_{max}$ and $p_{min}$ be the maximum and minimum values of $\vec{p}$ respectively. Let $m_{max}$ be the maximum value of $\vec{m}$. Then we have the following 
\begin{align}
        T_k(\vec{m},\vec{p})  & \leq  \frac{p_{max}}{p_{min}} T_k(m_{max},1) = & & \\&  \frac{p_{max}}{p_{min}} k( \log(k) +  (m_{max} - 1)\log(\log(k)) + o(1))  & &
\end{align}

\end{theorem}
\section{Problem Formulation}
\label{Sec:problem}
\subsection{Setting}
We consider the same setting as \cite{Waterfilling} adapted for online query answering. 

There are $k$ analysts each with an associated positive privacy budget $s_1, s_2 \dots s_k$ such that $\sum_{i=0}^k s_i = \epsilon$. These weights denote the shares of the overall privacy budget $\epsilon$ to which each analyst is entitled. We denote the privacy budget entitled to a collection of analysts $\gS \subseteq \{1, ..., k\}$ as $\epsilon_{\gS} = \sum_{i \in S} s_i$. \par 
A Differentially Private Multi-Analyst online mechanism takes in as input a sequence of \new{query, analyst} tuples $(q,i) \in \gQ \times \{1, \dots , k\}$ consisting of a linear counting query and the identifier for the analyst asking the query. We denote the sub-sequence of $\gQ$ containing only the queries asked by analyst $i$ as $\gQ^i$ and the sub-sequence of $\gQ$ containing queries from analysts in the set $\gS$ as $\gQ^{\gS}$ . Likewise we denote the $ith$ query in the query sequence $\gQ_i$. The data curator must answer each query one at a time before the next query is revealed. \par 
\new{Bun et al.  \cite{OnlineOffline} show that answering queries in the online setting is strictly more difficult than in the offline setting since the queries are not known in advance. This leads to classes of queries that can be answered efficiently under differential privacy in the offline case but not in the online case. Mechanisms designed for the offline case do not always translate straightforwardly to the online case. We will show that the additional constraints of the sharing incentive causes an even greater separation between the offline and online multi-analyst settings.
There are additional complexities that arise when considering multiple analysts in the online setting. In particular, we will show that the order in which queries are received can affect the distribution of error across queries. This is a unique property of the online setting which makes it difficult for a mechanism to share privacy across analysts budget while still ensuring the sharing incentive.}
\subsection{Desiderata}
The utility metrics used in online QA systems (number of sufficiently accurately answered queries) are often different from those used in offline QA systems (total mean squared error). In this work, we define utility as the number of queries an analyst can answer with error under some pre-defined threshold $\alpha$. Of course, the correct notion of utility may be context dependent, so we define the desiderata with respect to an arbitrary utility function $U_i$ representing the utility for analyst $i$. Given this, we adapt the multi-analyst desiderata from~\cite{Waterfilling} to the online setting. At a high level, these desiderata are intended to ensure that each of the multiple analysts receives high utility, as opposed to merely optimizing for total or average utility, potentially at the expense of some analysts seeing poor performance.
\par 
A natural baseline for the utility a given analyst should demand is that which they would expect if their queries were answered independently with their share of the privacy budget. Any multi-analyst differentially private mechanism should incentivize a rational agent to participate in the collective mechanism by guaranteeing that they will receive at least as much utility as they would expect in the independent case. Any mechanism which satisfies this requirement is said to satisfy the \textit{sharing incentive}, more formally as follows.
\begin{definition}[Sharing Incentive \cite{Waterfilling}] \label{SharingIncentive}
A mechanism $\gM$ satisfies the sharing incentive if for any collection of analysts $\gS$, any analyst $i \in S$ and all query sequences,
\begin{equation*}
\mathbb{E}[U_i(\gM,\gQ^{\gS},\epsilon_{\gS})] \geq 
\mathbb{E}[U_i(\gM,\gQ^i,s_i)]
\end{equation*}
\end{definition}
\new{Where $\mathbb{E}$ is the expectation over the randomness of the mechanism.} The sharing incentive only provides a baseline of comparison to utility in the independent case. A stronger guarantee would be that the addition of an analyst (along with their privacy budget) to any collective should never decrease the utility of any analyst in the collective. This property is called \textit{analyst monotonicity}. 

\begin{definition}[Analyst Monotonicity \cite{Waterfilling}]\label{def:monotonic}
A mechanism $\gM$ is analyst monotonic if for  all collections of analysts $\gS$, any two analysts $i, j \in \gS$ for all  query sequences. \\ 
\begin{equation*}
   \mathbb{E} [U_i(\gM, \gQ^{\gS}, \epsilon_{\gS})] \geq \mathbb{E}[ U_i(\gM, \gQ^{\gS \setminus j}, \epsilon_{\gS \setminus j})]
\end{equation*} 
\end{definition}

It is straightforward to show that analyst monotonicity implies sharing incentive by induction. The converse, however, is not true; analyst monotonicity is a stronger guarantee in this sense. For example, suppose you know the utilities guaranteed by the sharing incentive to analysts from their independent cases. You could satisfy the sharing incentive by optimizing for total number of queries answered subject to utility constraints for all of the analysts, but such a scheme would not ensure analyst monotonicity in general.

Note that checking for violation of analyst monotonicity empirically is intractable in general, as it requires quantifying over all possible subsets of analysts. The relaxation of analyst monotonicity called \textit{non-interference} provides the same guarantee but only for the collective consisting of all analysts. In other words, the weaker guarantee of non-interference is that the addition of the \textit{last} analyst to the collective does not decrease the utility of any other analyst. We will argue theoretically that our algorithms satisfy the stronger property of analyst monotonicity or the weaker property of sharing incentive, but we include the definition of non-interference because we measure the empirical interference in our experiments. 

\begin{definition}[Non-Interference \cite{Waterfilling}]\label{def:non-int}
A mechanism $\gM$ satisfies non-interference if for any two analysts $i,j \in \gS$ where $i \neq j$, and for all  query sequences, \\ \begin{equation*}
   \mathbb{E} [U_i(\gM, \gQ, \epsilon)] \geq \mathbb{E}[ U_i(\gM, \gQ \setminus \gQ^j, \epsilon - s_j)]
\end{equation*} 
\end{definition}

Finally, any multi-analyst mechanism should be able to adapt and efficiently answer any valid query. We say that a multi-analyst mechanism is \textit{Adaptive} if the sequence of outputs given is a direct function of its input queries and is not pre-determined. This ensures that in addition to satisfying the desiderata above mechanisms are efficient and non-trivial.

\subsection{Independent Mechanisms}
Prior work \cite{Waterfilling} has shown that one can satisfy all three desiderata by dividing the privacy budget and running an instance of a single analyst differentially private mechanism (with their associated privacy budget). These \textit{independent mechanisms} however are highly inefficient as they do not allow analysts with similar queries to share either privacy budget or query answers. We therefore, take these independent mechanisms as an appropriate baseline to which we can compare our new mechanisms.
\subsection{Problem Statement}
The goal of this work is to design online multi-analyst differentially private mechanisms that answer (possibly large) query sequences submitted by multiple analysts while satisfying the three desiderata.

\begin{problem}
Given any sequence of query, analyst tuples $\gQ$ on a database $D$ with positive weights $s_1, \ldots, s_k$ s.t. $s_1+ \ldots + s_k = \epsilon$, design an adaptive mechanism $\mathcal{M}$ such that: 
\begin{itemize}
    \item $\mathcal{M}$ satisfies differential privacy,
    \item  $\mathcal{M}$ satisfies sharing incentive (\cref{SharingIncentive}), analyst monotonicity (\cref{def:monotonic}), and non-interference  (\cref{def:non-int}),
    \item and $\mathcal{M}$ answers as many queries accurately as possible.
\end{itemize}
\end{problem}
\section{Motivating Experiments}
\label{sec:motivation}
Here we demonstrate that \new{both classic mechanisms such as the Laplace mechanism as well as state of the art mechanisms such as Private Multiplicative Weights}  fail to satisfy the sharing incentive and non-interference. The experiments highlight a key difference between the online and offline problem, namely that the order in which queries are answered can have a significant impact on the utility of the agents.

Consider the case with two analysts, Alice and Bob, each with an identical query sequence on disjoint halves of the dataset and an equal share of half the privacy budget $\epsilon = 1$. 
In order to generate the query sequences, we let $p \in [0.5, 1]$ determine the probability that Alice's query gets asked at each time step. At $p=0.5$, Alice and Bob have an equal chance of their queries being answered; the resulting joint query stream is uniformly distributed between their queries. As $p$ increases at each time step the probability that the query will belong to Alice increases. When $p \rightarrow 1$, all of Alice's queries are answered first followed by Bob's queries.

\new {In the following experiments, the online mechanisms used were the Laplace mechanism and Private Multiplicative Weights (PMW) \cite{PMW}.} We use the practical census database provided in \cite{HDMM}. This database contains information about population migration by age and is of dimension $|x|= 86$ representing ages from $0$ to $85$. 

For the following figures, we generated a randomized workload, containing point and range queries, 1000 times, and ran PMW for each  value of p, in $ [0.5, 0.6, 0.7, 0.8, 0.9, 1]$. We examined the percent of queries answered with error under the threshold $\alpha= 0.01$ for each analyst's workload (Fig 1a) and the  Ratio Error (Fig 1b), a metric first introduced in \cite{Waterfilling}. The ratio error measures the ratio between the number of queries answered with error under $\alpha$ in the joint case and the number of queries answered with error under $\alpha$ in the independent case for each analyst.
Values above 1 signify a violation of the sharing incentive with larger values signifying a larger violation. The shaded regions of the plot represent the 90\% confidence intervals.

\begin{figure}[h]
\begin{subfigure}[b]{0.45\linewidth}
\includegraphics[width=\linewidth]{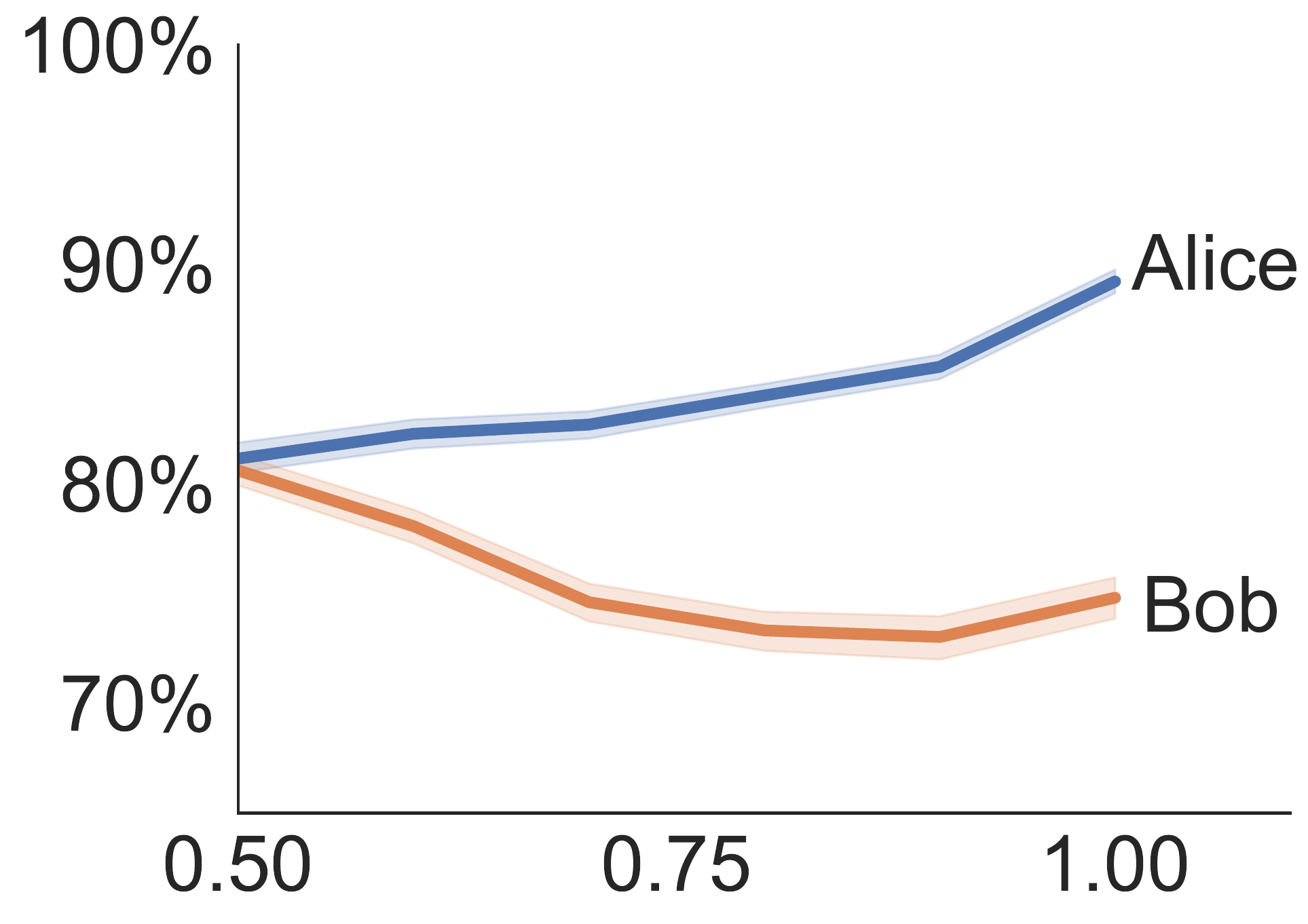}
 \label{fig:f1}
\end{subfigure}
\begin{subfigure}[b]{0.45\linewidth}
\includegraphics[width=\linewidth]{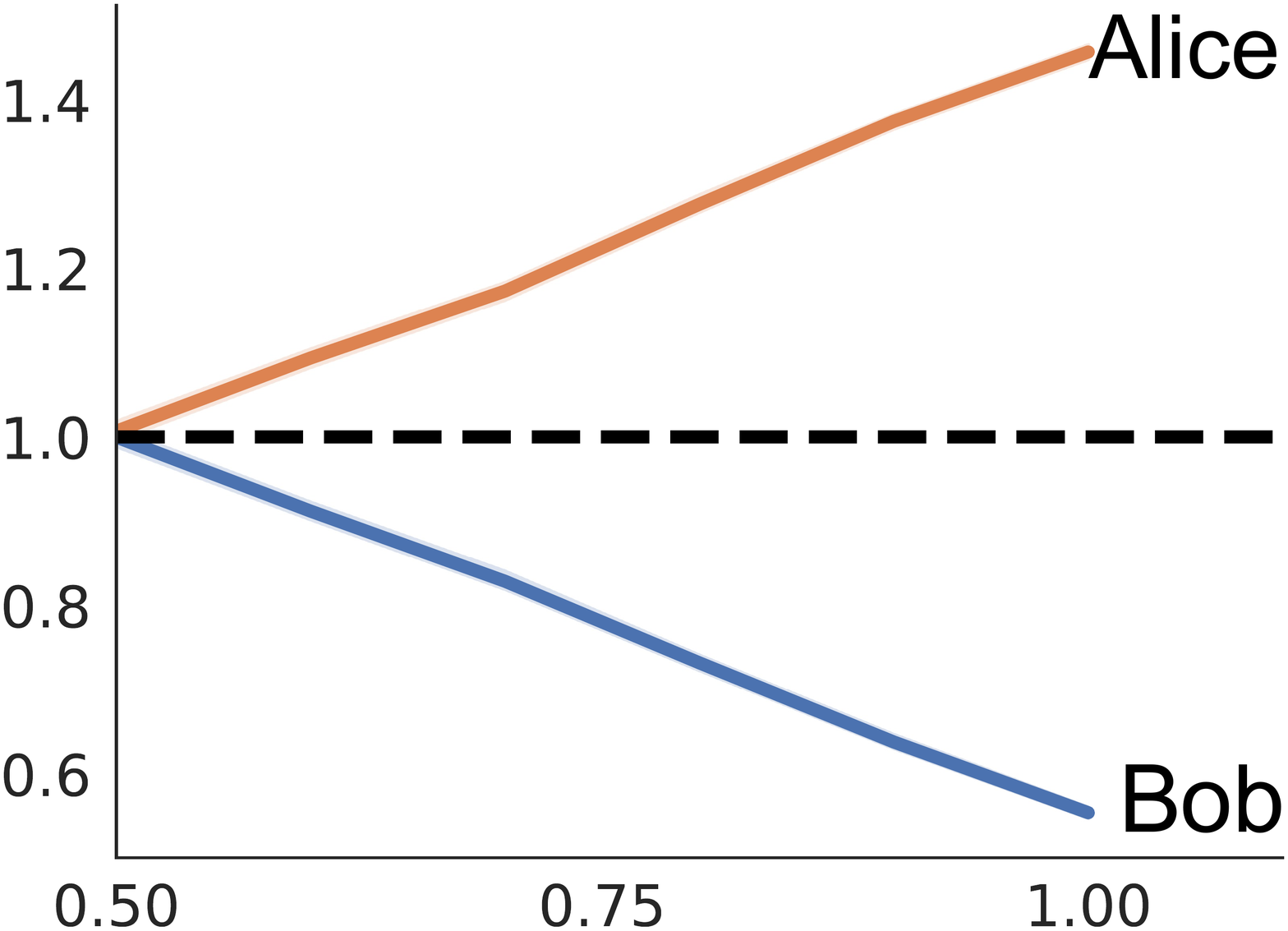}
 \label{fig:f2}
\end{subfigure}
\begin{subfigure}[b]{0.45\linewidth}
\includegraphics[width=\linewidth]{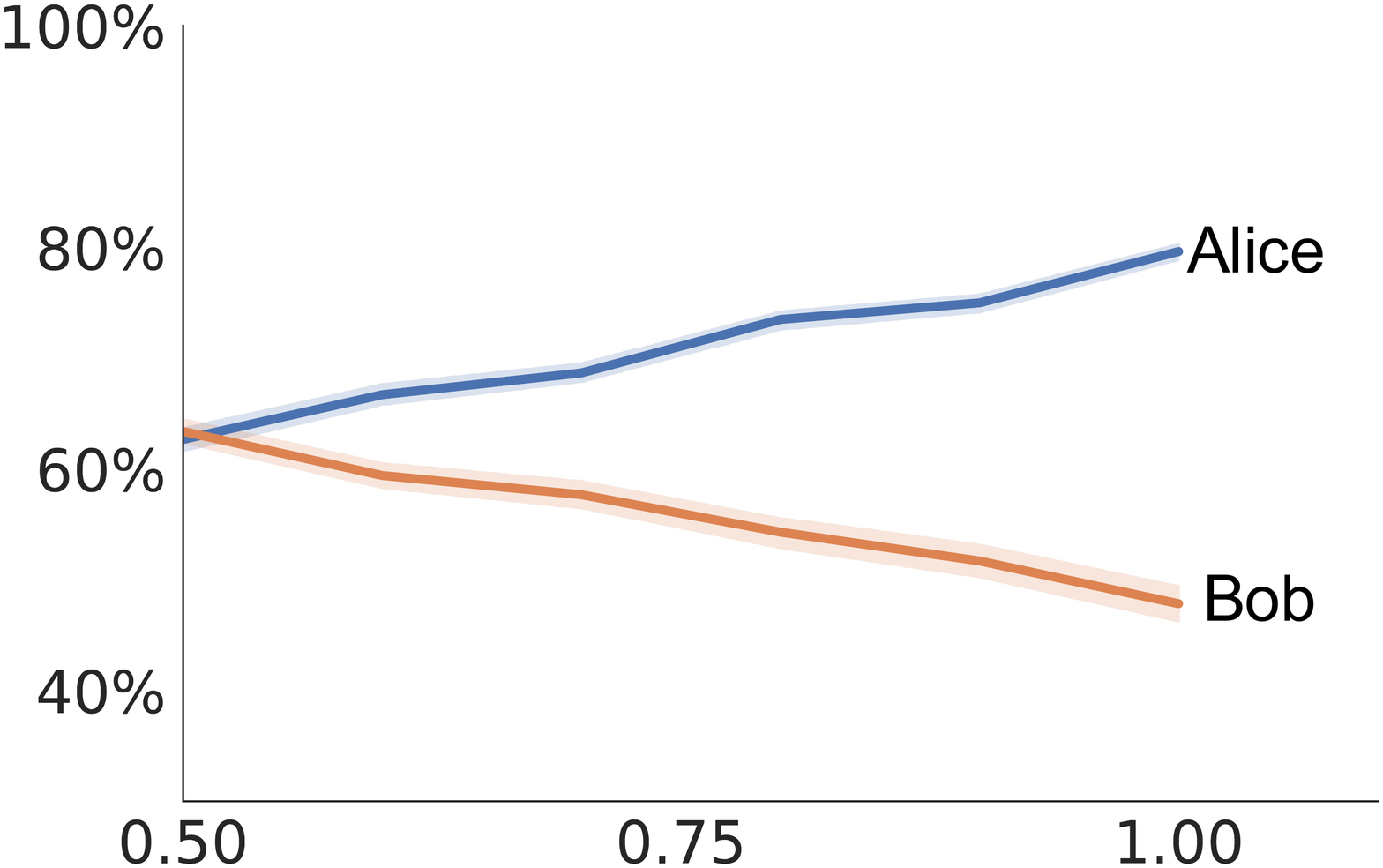}
 \caption{ Queries Answered}
 \label{fig:f1lap}
\end{subfigure}
\begin{subfigure}[b]{0.45\linewidth}
\includegraphics[width=\linewidth]{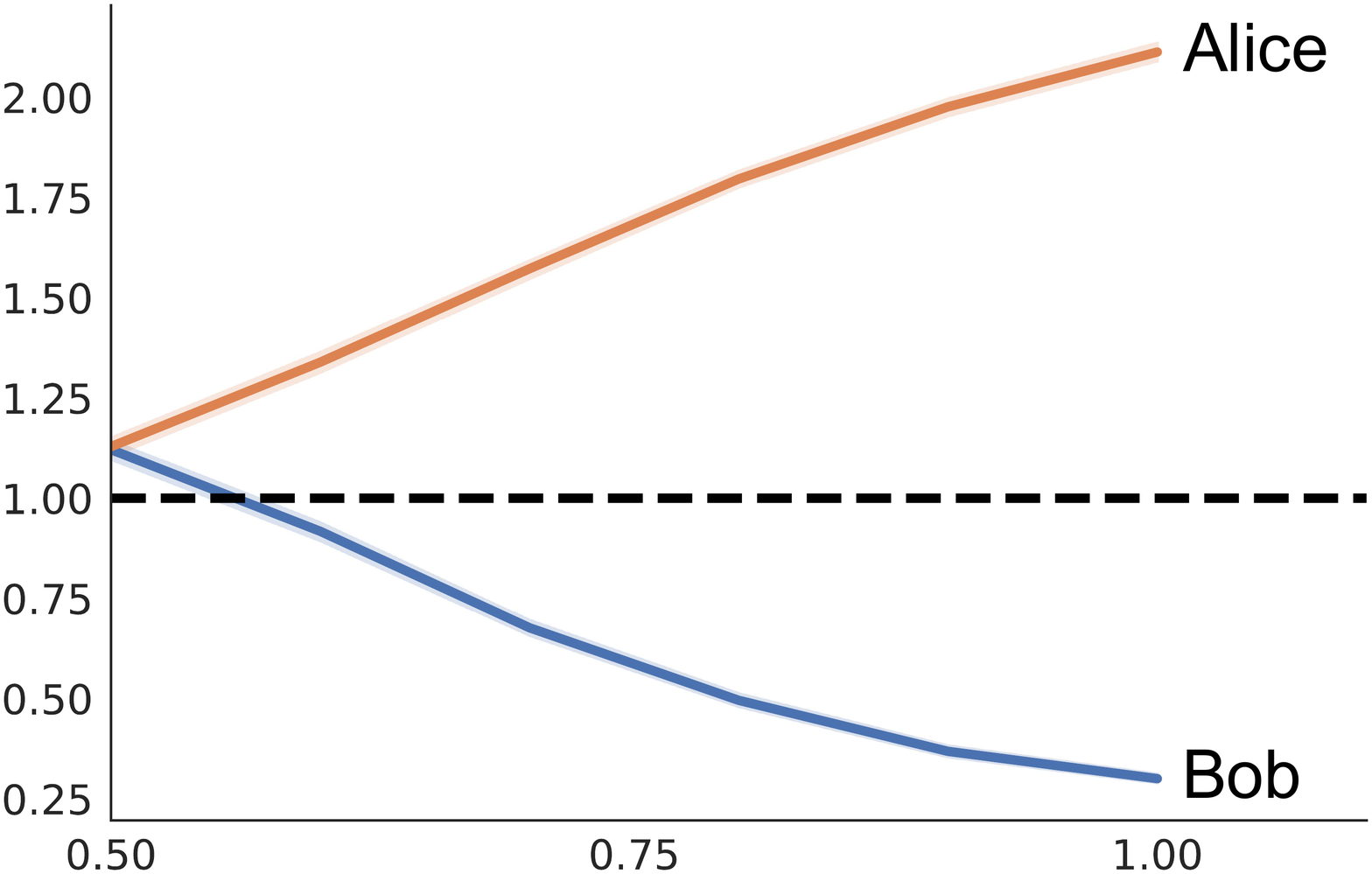}
\caption{Max Ratio Error} \label{fig:f2lap}
\end{subfigure}
\caption{Empirical Measures for PMW (above) and Laplace (below). Values of Utility (Left) and Sharing Penalty (Right) are shown with varying values of p (x-axis). }
\end{figure}

\new{The following results are consistent across both mechanisms.} In \cref{fig:f1lap}, we show that utility disparities between analysts grow as $p$ increases. Alice and Bob have the same utility at $p = 0.5$ but as p increases, Alice's utility increases and Bob's utility decreases. Even in a situation with only 2 analysts with very similar queries, the ordering of the queries can significantly impact the utility of each analyst. In a system designed without multi-analyst desiderata in mind, those analysts who ask their queries early may receive a large benefit for doing so. This could create an incentive for analysts in such systems to "race" to submit queries first to ensure higher utility.

\cref{fig:f2lap}, demonstrates that this increased disparity can cause a mechanism to fail to satisfy the sharing incentive. Here we test an instance where Alice and Bob have identical query sequences on disjoint partitions of the database (e.g., males and females).  Although they start with the same max ratio error when p = 0.5, the disparity grows as p grows. This increased disparity causes a failure to satisfy the sharing incentive as, in the worst case, Bob receives $40\%$ more error than he would have in the independent case.
\section{Limitations in the Online Setting}
\label{Sec:limitations}
\begin{figure}[t]
   \centering
    \includegraphics[width= 0.9 \linewidth]{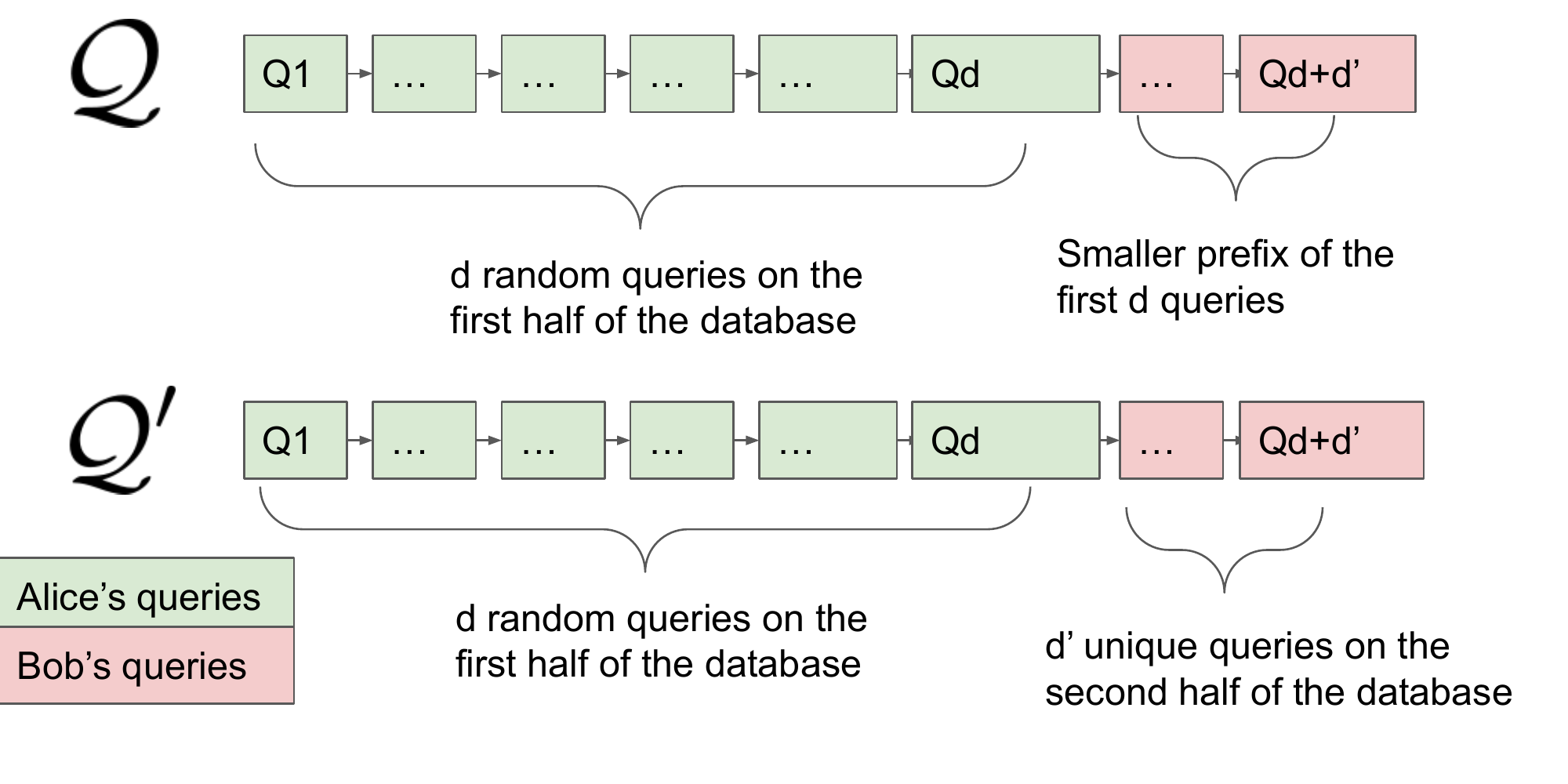}
    \caption{Visualization of the two sequences $\gQ$ and $\gQ'$}
    \label{fig:seperation}
  \end{figure}
  
Here we will introduce a fundamental performance cost for online mechanisms satisfying the sharing incentive. Specifically, we prove Theorem~\ref{Thrm:SharingIncentiveUpper}, which shows that there is an upper bound to the number of queries that can be accurately answered while satisfying the sharing incentive, and that the number of queries decreases with $k$, the number of analysts. \par
The argument takes advantage of the online nature of the mechanism. We provide two sequences $\gQ$ and $\gQ'$ which contain the same (significantly long) prefix. We will argue that in order to satisfy the sharing incentive and answer all the queries with expected error below an arbitrary threshold $\alpha$, the mechanism would need to have different behavior during the prefix in both $\gQ$ and $\gQ'$. Since the prefixes are identical no online mechanism can differentiate between the two. Consequently, a mechanism must choose between answering many queries consistently and satisfying the sharing incentive.  \par 

In order to establish the result, we rely on a few previous theorems. First, we require the lower bound on error for any pure $(\epsilon , 0)$-differentially private mechanism. 
\begin{theorem}[Differential Privacy Lower Bound \cite{hardt2009geometry}] \label{Thrm:Geometry}
Any mechanism $M$ which satisfies differential privacy must suffer at least $\Omega(|\gQ|/\epsilon) \cdot  \min \left( \sqrt{\log(n/|\gQ|)}, \sqrt{|\gQ|} \right) $ expected error 
\end{theorem}

We then need to establish the existence of a mechanism which can answer a large enough number of queries with a set privacy budget. 
\begin{theorem}\label{Pure_existance_1}
There exists an $\epsilon$ differentially private query answering mechanism which can answer $|\gQ|$ queries, each with error less than or equal to $\frac{|\gQ| \sqrt{|\gQ|}}{\epsilon}$
\end{theorem}
\begin{proof}
\new{
By splitting the privacy budget evenly across each of the $|\gQ|$ queries then applying the Laplace Mechanism to answer each query you can answer $|\gQ|$ queries with expected error exactly $\frac{|\gQ| \sqrt{|\gQ|}}{\epsilon}$ \cite{LM}. By \cref{thrm:composoition} this satisfies $\epsilon$-differential privacy.}
\end{proof}

 \begin{theorem}[Sharing Incentive Query Limit]
\label{Thrm:SharingIncentiveUpper}
\new{For all online multi-analyst $\epsilon$-DP query answering mechanisms $M$, for any number of analysts $k$, any database $x$ of dimension $|x|  \geq 2k$ and $n$ sufficiently large there exists shares of privacy budgets $(s_1, \dots ,s_k)$ and a query sequence $\gQ$ of size $O(\left(\alpha\epsilon /k\right)^{2/3})$ such that if mechanism $\gM$ can answer all the queries in $\gQ$ with error less than threshold $\alpha$ then there exists an alternative workload $\gQ'$ in which $\gM$ must violate the sharing incentive.} 
\end{theorem}

\begin{proof}
We begin by considering the case where there are two analysts, Alice and Bob and will expand to $k$ analysts afterward. Each analyst is entitled to $\epsilon/2$ of the privacy budget. We will construct two query sequences $\gQ$ and $\gQ'$ which share a large prefix such that if a mechanism spends the entire privacy budget answering the prefix it can answer all the queries in $\gQ$ but if it spends more than $\epsilon/2$ budget answering the prefix, it will violate the sharing incentive in $\gQ'$.\par

We start by constructing sequence $\gQ$. First we partition the database \new{on a single predicate} into two disjoint halves (e.g., males and females). Alice will ask queries on the first half of the database. Bob will also ask his queries on the first half in $\gQ$. He will ask his queries on the second half in $\gQ'$. Alice asks $d$ random queries on the first half of the database where $d$ is the largest integer such that $\frac{d \sqrt{d}}{\epsilon} < \alpha < \frac{2d \sqrt{d}}{\epsilon}$ . By \cref{Thrm:Geometry}  Alice's queries cannot be answered by any mechanism using only $\frac{\epsilon}{2}$ of the privacy budget \new{but can with the entire privacy budget}. Bob will then ask the first $d'$ queries from Alice's queries where $d'$ is the largest integer such that  $\frac{2d' \sqrt{d'}}{\epsilon} < \alpha$. In this case, all of Bob's queries are identical to Alice's queries and as such her query answers can be reused to answer Bob's. \par 

Now we will construct the alternative sequence $\gQ'$. Alice will ask the same queries as in $\gQ$, this will serve as the identical prefix. In this case, Bob will ask a different set of queries. Bob will ask $d'$ random distinct queries from the second half of the database where $d'$ is the largest integer such that $\frac{2d' \sqrt{d'}}{\epsilon} < \alpha <\frac{(d+d') \sqrt{(d+d')}}{\epsilon}$. In this case, since Bob's queries are from the second half of the database Alice's queries cannot be used to aid in answering Bob's queries. \par

Since Bob's queries in $\gQ$ are copies of Alice's queries, all the queries in $\gQ$ can be answered by answering only Alice's queries using the entire privacy budget and reusing her answers to answer Bob's queries. By \cref{Pure_existance_1} the Laplace mechanism can answer all of Alice's queries under the threshold $\alpha$. Since $ \alpha < \frac{2d \sqrt{d}}{\epsilon}$,  by \cref{Thrm:Geometry}, there is no differentially private mechanism that can answer Alice's queries while only using $\epsilon/2$ of the privacy budget. In order to answer all of Alice's queries some of Bob's budget must be used. Likewise in $\gQ'$ since $\frac{(d+d') \sqrt{(d+d')}}{\epsilon} > \alpha$ there exists no $\epsilon$-differentially private mechanism that can answer all of Alice's queries and all of Bob's queries under the threshold. However, no online mechanism can distinguish between $\gQ$ and $\gQ'$ before Bob's queries and thus must have the same behavior on both sequences up to this point. Since Alice's queries in both are sufficiently large the mechanism must decide on either answering all of Alice's queries or saving privacy budget for Bob.  If mechanism $\gM$ can answer all the queries in $\gQ$ it must use  Bob's share of the privacy budget prior to Bob's queries being answered. In $\gQ$ this is fine as Bob's queries consist of a prefix of Alice's queries which can be reused, however in $\gQ'$, since $d'$ is the largest integer such that$\frac{2d' \sqrt{d'}}{\epsilon} < \alpha$ by \cref{Thrm:Geometry}, if any of Bob's privacy budget is used prior to his queries there exists no mechanism that can answer all of his queries. 
Since Bob's share of the privacy budget is sufficient for mechanism $\gM$ to answer all of his queries in the independent case, this is a violation of the sharing incentive.
Alternatively, if mechanism $\gM$ does satisfy the sharing incentive it cannot answer all of Alice's queries in either $\gQ$ or $\gQ'$ since it cannot distinguish between $\gQ$ and $\gQ'$ prior to Bob's queries. \par

We can extend this example to the case of $k$ analysts to get the
linear separation between a general online mechanism and one that satisfies the sharing incentive. We consider $k$ analysts each with $\frac{\epsilon}{k}$ of the privacy budget and split the database into $k$ equal parts. The first analyst will ask $d$ queries where $d$ is the largest integer such that $\frac{d \sqrt{d}}{\epsilon} < \alpha < \frac{kd \sqrt{d}}{\epsilon}$.

Each subsequent analyst will each ask a  set of $d'$  queries where $d'$ is the largest integer such that $\frac{kd' \sqrt{d'}}{\epsilon} < \alpha$ and $\frac{(d+(k-1)d') \sqrt{(d+(k-1)d')}}{\epsilon}$ $ > \alpha$ .  In $\gQ$ each of those analysts will ask prefixes of the first analyst's queries and in $\gQ'$ they will each ask queries from their unique partition of the database.

\new{By \cref{Thrm:Geometry} no mechanism can answer more than$\left(\alpha\epsilon /k\right)^{2/3}$ queries using with error under $\alpha$ using $\frac{\epsilon}{k}$ of the privacy budget.  }
As such if the mechanism answers any more than $\left(\alpha\epsilon /k\right)^{2/3}$ of the first analyst's queries it must take from at least one of the other subsequent analysts resulting in a violation of the sharing incentive.

Therefore any mechanism  that satisfies the sharing incentive in all cases can answer at most $\left(\alpha\epsilon /k\right)^{2/3}$ queries for any agent when the privacy budget is distributed equally. Now consider $\gQ''$ which shares the same prefix as $\gQ$ and $\gQ'$. In this case, each analyst other than the first only asks one query. Since this query sequence contains the same large prefix it must have the same behavior as in the previous two cases. Thus despite being asked many queries a mechanism that satisfies the sharing incentive can only answer $ \left(\alpha\epsilon /k\right)^{2/3} + k-1$ queries in total.
\end{proof}
This shows that any online mechanism can choose between answering a large number of queries in all cases or satisfying the sharing incentive but not both. 
Compare this to the number of queries that can be answered by the simple Laplace mechanism with sequential composition. The Laplace mechanism can answer an arbitrary  $\left(\alpha\epsilon\right)^{2/3}$ queries under the threshold $\alpha$. This means that even the Laplace mechanism with sequential composition can answer too many queries, in any sequence of queries, to satisfy the sharing incentive. Any mechanism which satisfies the sharing incentive cannot perform any better in all cases than the independent version of the Laplace mechanism which divides the privacy budget equally k ways. \par
We note that this result is unique to the online setting where queries must be answered one at a time and in a fixed order. In the offline setting where the entire workload is known one can create an alternative "strategy workload" to answer which can be used to reconstruct each analysts' queries efficiently.
Often online mechanisms are analyzed through the random order model \cite{RandomOrder,OnlineRandomInputModel,BugetedRandomInputModel} where the queries appear in random order instead of an adversarial one. However, in our case ordering the queries randomly doesn't circumvent the upper bound of \cref{Thrm:SharingIncentiveUpper} since, \new{in the setting above}, the first analyst asks the vast majority of queries. As the number of analysts and privacy budget grows large the first analyst's queries become the overwhelming majority of queries and as such the adversarial case will still happen with high probability. \par 
In the following sections, we will present two solutions for online multi-analyst differential privacy. The first solution in \cref{Sec:CacheAndReconstruct} will be an analyst monotonic mechanism for any sequence of queries and any distribution of privacy budget. As such this solution will be subject to the upper bound of \cref{Thrm:SharingIncentiveUpper}. In \cref{Sec:RandomOrdering} we will introduce a method to circumvent \cref{Thrm:SharingIncentiveUpper} by restricting the order in which queries can appear. From there we will show that any sufficiently efficient online mechanism can be made to satisfy the sharing incentive by restricting the order in which analyst queries are answered.
\section{Cache and Reconstruct}
\label{Sec:CacheAndReconstruct}

Here we introduce Seeded Cache and Reconstruct (Algorithm~\ref{Algo:SCR}), an online mechanism for answering linear queries which satisfies all the desiderata  for any sequence of queries in any order.
The seeded cache and reconstruct mechanism works by initially generating a cache of answered queries and then using that cache to reconstruct other queries once privacy budget has been expended. The mechanism has three phases \new{which each analyst goes through asynchronously based on their remaining privacy budget}:
\begin{itemize}
    \item \new{The first phase happens during initialization.} Each analyst donates a fixed fraction of their privacy budget to answer a basis set of queries using all the donated budget. As the number of analysts increases the quality of the seed is increased as more privacy budget is donated to it. An example of such a basis is a histogram of counts over all unique values in the database's universe.
    \item \new{The second phase begins after initialization and ends once the analyst runs out of privacy budget. In this phase, the analysts ask additional queries. If the query is already in the cache the answer is reused. If the query is not in the cache it is answered using the Laplace mechanism and then added to the cache.}
    \item \new{The third phase begins once an analyst's privacy budget is expended.} In this phase, they can use the entire cache to reconstruct their remaining queries by using the Matrix Mechanism \cite{Matrix} reconstruction step. Since the basis was generated in the first phase it is possible to reconstruct any query from the cached queries but those reconstructions may lead to answers with high error.
\end{itemize}
\begin{lemma}
Seeded Cache and Reconstruct is analyst monotonic.
\end{lemma}
\begin{proof}
Consider a collective $S \subseteq \{1, ..., k\}$ and analysts $i, j \in S$. $\gQ^{\gS \setminus j}$ is the subsequence of $\gQ$ restricted to queries labeled for analysts in $\gS \setminus j$. Consider some $q \in Q^{i}$ such that Seeded Cache and Reconstruct answers $q$ with error at most $\alpha$ when run on $\gQ^{\gS \setminus j}$ with privacy budget $\epsilon_{\gS \setminus j}$. We will show that Seeded Cache and Reconstruct also answers $q$ with error at most $\alpha$ when run on $\gQ^{\gS}$ with privacy budget $\epsilon_{\gS}$. The argument follows by cases for how $q$ is answered by Seeded Cache and Reconstruct in the first instance (without $j$ in the collective).

First, suppose $q$ was answered directly by using privacy budget from analyst $i$ in the first instance. Then $q$ will still be answered as long as $i$ has sufficient privacy budget when asking $q$ in the second instance. This must be the case: For any prior query on which $i$ spends privacy budget in the first instance, either $i$ still spends the same privacy budget on the query in the second instance, or the query was asked by $j$ and cached, in which case $i$ will have more privacy budget in the second instance. In other words, individual analyst privacy budgets at a given query in their sequence are monotone non-decreasing with respect to the addition of an analyst to the collective.

Second, suppose the answer to $q$ was in cache in the first instance: Then some other analyst $k \neq i \in \gS \setminus j$ spent budget to answer $q$ prior to $i$'s query in $\gQ^{\gS \setminus j}$. Because answering directly always takes precedence over reconstruction when possible, $k$ will also spend budget to answer $q$ prior to $i$'s query in $\gQ^{\gS}$ unless $j$ has already answered and cached $q$ previously. Either way, $q$ will still be cached when $i$ asks the query in $\gQ^{\gS}$. Note that $k$ must still have enough budget to afford $q$ in $\gQ^{\gS}$ by the budget monotonicity condition argued in the first case. In other words, the cache contents are also monotone with respect to adding an analyst to the collective.

Third and finally, suppose $q$ was answered by reconstruction using the Matrix Mechanism reconstruction step in the first instance (with expected error at most $\alpha$). Suppose for a contradiction Seeded Cache and Reconstruct returned a reconstructed answer to $q$ with expected error greater than $\alpha$ in the second instance with $j$ in the collective. But we have already argued in the second case that the cache contents are monotone with respect to adding an analyst to the collective, and \cref{lemma:rows} states that any such reconstruction on a superset of the first instance's cache can only have lower expected error. Likewise, the seed quality monotonically increases with the number of analysts. \cref{lemma:diagonal} states that any reconstruction with identical queries of higher quality can only have lower expected error.  This establishes the contradiction in the third case.

Thus, the queries of analyst $i$ answered in the first instance (without $j$ in the collective) with expected error at most $\alpha$ is a subset of the queries answered with expected error at most $\alpha$ in the second instance (with $j$ in the collective). As the analyst's utility is the number of such queries, it follows that $i$'s utility in the first instance is no more than $i$'s utility with $j$ in the collective.

\end{proof}
\SetKw{KwBy}{by}
\begin{algorithm}[h]
\SetAlgoLined

\caption{Seeded Cache and Reconstruct}
\label{Algo:SCR}

\SetKwInOut{Input}{input}\SetKwInOut{Output}{output}
\Input{Sequence of queries with associated analyst $\gQ \leftarrow (q_1,a_i) \dots (q_{|\gQ|},a_i) $,\\ 
Vector of $k$ shares  $\vec{S} \leftarrow \{s_1, s_2 \dots s_k\ \} $ , \\
Data vector $\mathbf{x}$,\\
privacy budget $\epsilon$,\\
Threshold $\alpha$, \\
Fraction of budget for seed $\gamma$, \\
Basis of queries to be generated $\mB$,\\
Privacy budget per query $\lambda$} 
\Output{Sequence of Query Answers}
\vspace{3 mm}
\nonl \textbf{\new{Phase 1 (Mechanism Initialization)}} \\
 Seed cache $\mC$ with all queries of $B$ with $\gamma\epsilon$ privacy budget  \\
 Create a vector of remaining privacy budget $\vec{\epsilon} =  (1-\gamma)\epsilon \vec{S} $ \\
 \For{$i\gets1$ \KwTo $|\gQ|$ \KwBy $1$}{
    \vspace{3 mm} \nonl \textbf{\new{Phase 2 (Using Privacy Budget)}} \\
    \If{$\vec{\epsilon}_i \geq \lambda$}{

    \If{$\gQ_i \in \mC$}{
      \Return Query answer from $\gC$
      }
      \Else{
        Create noisy query answer $\hat{\gQ_i}$ with Laplace Mechanism using privacy budget $\lambda$ \\
        $\vec{\epsilon}_i  \leftarrow \vec{\epsilon}_i  - \lambda $ \\
        Add $\hat{\gQ_i}$ to $\gC$ \\ 
        \Return $\hat{\gQ_i}$
        
    } }
    \vspace{3 mm}\nonl \textbf{\new{Phase 3 (Reconstructing Queries)}} \\
    \If{$\vec{\epsilon}_i < \lambda $}{
        \Return Matrix Mechanism reconstruction of $\gQ_i$ using $\gC$
        }
        
    }
 
\end{algorithm}

\subsection{Utility and Trade-offs}
Each component to seeded cache and reconstruct adds additional utility to the mechanism while retaining analyst monotonicity. The reconstruction allows for additional queries to be answered once an analyst's privacy budget is expended. The full rank seeding at the beginning ensures that the matrix mechanism reconstruct step \cite{Matrix} can reconstruct any linear query and not just those supported by the queries that are added to the cache. This ensures that even when all analysts have expended their privacy budget any query may be answered (though not necessarily with error below the $\alpha$ threshold).

Seeded cache and reconstruct always satisfies Analyst Monotonicity, \new{regardless of the seed and parameter choice, } (and thus the weaker notion of Sharing Incentive) but the choice of seed can greatly impact the error of reconstructed queries. Seeded cache and reconstruct is at its best when one has some prior knowledge of the likely kinds of queries to be asked. In this case, the data curator can choose a specific seed that performs well on those queries. Past work \cite{Hb,AHP,Ding2011, Hay2010} has shown that one can design a particular workload that performs well on a family of queries. For example the hierarchical mechanism \cite{Hb} performs especially well on long range queries. A data curator who knows that range queries will be asked frequently may use the queries from the hierarchical mechanism as a basis in that case. Even in cases where there is no known optimal basis for a particular family of queries the data curator can use the Matrix Mechanism \cite{Matrix} to generate a strong basis. \par

One may note that seeded cache and reconstruct is restrictive in how it answers queries: the algorithm never attempts to reconstruct the answer to a new query if there is privacy budget available to answer it via Laplace. There are times when this seems to imply an unnecessary waste of privacy budget. We argue that this restriction is necessary to ensure analyst monotonicity, specifically the sharing incentive.

For example, consider the case where there are two analysts Alice and Bob. They each have equal shares of the privacy budget. Assume that any query reconstructed with at most 2 queries which are answered using the privacy budget will have error under the threshold $\alpha$. Alice asks her queries first and asks all the point queries (range queries of length 1). Bob asks all the length 2 range queries followed by all the length 4 range queries. In the independent case Bob will answer all of his length 2 range queries by expending his privacy budget and will reconstruct his length 4 range queries using those answers. In the joint case however Alice will answer all her queries using her privacy budget, then Bob will reconstruct the size 2 range queries from Alice's cached queries. When he asks the size 4 range queries he will not be able to reconstruct them all from the queries in the cache and will fail in answering all of them with privacy budget directly. These conditions arise when the addition of new analysts results in drastically different behavior between the joint and independent cases. \par 

Since seeded cache and reconstruct satisfies the sharing incentive it is subject to \cref{Thrm:SharingIncentiveUpper}. Seeded cache and reconstruct performs particularly poorly in cases like the ones shown in \cref{Sec:limitations} where each analyst asks queries on disjoint sections of the database. In these cases, no two analysts have any queries that rely on the same information. As no analyst can use another analyst's query answers to help reconstruct their own. In this case, each analyst performs no better than if they had answered their queries independently. While these cases do exist we show in \cref{experiments} that these cases are infrequent and that cache and reconstruct typically outperforms even optimal independent mechanisms. \par 
Despite this, there is a cost that analyst monotonic mechanisms pay in terms of their ability to a large number of queries below $\alpha$ error. We see in \cref{experiments} that mechanisms that fail to satisfy any of the desiderata such as non-independent PMW outperform seeded cache and reconstruct in this sense.

\vspace{-1em}
\section{The Query Scheduler}
\label{Sec:RandomOrdering}

Seeded Cache and Reconstruct achieves all of our desiderata, and in a sense ``solves'' the problem of multi-analyst differential privacy for online query answering. However, it is subject to the fundamental upper bound of Theorem~\ref{Thrm:SharingIncentiveUpper} which implies that performance (measured in the total number of queries answered) has a polynomial dependence on $k$, the number of analysts. In system design contexts where privacy budget is an extremely precious resource, this may not be an acceptable trade-off. 

Instead, one might want to use a given state-of-the-art online query answering algorithm without compromising performance and simply modify it to satisfy the sharing incentive for multiple analysts. In this section, we describe how to complete such a generic reduction. Our intuition from Section~\ref{sec:motivation} is that the \textit{order} in which analysts ask queries is also of crucial importance to multi-analyst desiderata (in particular, it is better from an analyst's perspective to come earlier in the order). Whereas Seeded Cache and Reconstruct achieves analyst monotonicity by directly accounting for the privacy budget used by different analysts, we show that it is also possible to satisfy the sharing incentive  by constraining the order in which analysts ask queries. 

We introduce the \textbf{Query Scheduler}. The query scheduler takes in as input an existing online mechanism and the query sequence. At each time step, an analyst is chosen to answer their next query in the sequence. While seeded cache and reconstruct allocates privacy budget directly, the Query Scheduler instead ``allocates'' \textit{time}, by selecting at which times analysts can answer their queries. Note that this approach circumvents Theorem~\ref{Thrm:SharingIncentiveUpper} precisely by restricting the online input model. Importantly, we do not constrain the set of possible query sequences, only the set of possible analyst identifier sequences. Practically speaking, this has the effect of modifying existing algorithms to satisfy the sharing incentive at the cost of occasionally ``stalling'' when the next scheduled analyst does not have a query ready. 
 
Below we introduce two methods for allocating time across analysts. The \textbf{round-robin Scheduler} enforces that queries be asked in a round-robin fashion where every analyst must answer a query before any other analyst can ask another. The \textbf{Randomized Scheduler} instead randomly selects an analyst at each time step to answer their queries.

\subsection{Round-robin Scheduler}

The round-robin scheduler selects analysts one at a time in a fixed rotation. This ensures that every analyst is given a chance to answer a query before any analyst gets another.
 Consider applying the round-robin scheduler to the adversarial query sequence from \cref{Sec:limitations} where Alice asks all her queries first followed by a single query from each other analyst. The round-robin scheduler will ensure that each analyst receives an opportunity to ask their queries before Alice expends all of the privacy budget.  As a result the scheduler stalls after Alice's first query waiting for the queries from the other analysts. In this case, Alice must wait until the entire query sequence is complete before her second query is answered. The round-robin scheduler ensures that each analyst will have the opportunity to answer their queries before all the resources are consumed at the cost of additional wait time.
 As a result \cref{Thrm:RoundRobin} states that any mechanism which scales at least linearly with the privacy budget satisfies the sharing incentive.

\begin{theorem} [round-robin]
\label{Thrm:RoundRobin}
Let every analyst have an equal share of the privacy budget. When queries are asked in a round-robin  manner any mechanism $\gM$ which can answer at most $c$ queries with expected error under threshold $\alpha$ with privacy budget $\epsilon$ and can answer at least the first $k\cdot c $ queries under threshold $\alpha$ with privacy budget $k \cdot \epsilon$ satisfies the sharing incentive.
\end{theorem}

\begin{proof}
First, we note that as a property of the round-robin scheduler if there are $k$ analysts after answering $c\cdot k$ queries each analyst has asked exactly $c$ queries. Each analyst is entitled to an equal privacy budget $\frac{\epsilon}{k}$. Therefore when there are $k$ analysts participating there is a total of $ \epsilon$ privacy budget. By assumption mechanism $\gM$ can answer at most $c$ queries and at the end of $c$ queries each analyst has had $\frac{c}{k}$ queries answered. Let an additional analyst join the system. The total privacy budget is now $\epsilon + \frac{\epsilon}{k} = (1 + \frac{1}{k})\cdot \epsilon$. By assumption mechanism $\gM$ can answer at least $c + \frac{c}{k} = \frac{(k+1)c}{k}$ queries resulting in each analyst answering at least $\frac{(k+1)c}{k} \frac{1}{k+1} = \frac{c}{k}$ queries. Therefore the addition of any analyst can only ever improve the number of queries each analyst answers. 

\end{proof}

We can further adapt this for cases with non-equal shares of the privacy budget. Instead of doing strict round-robin ordering during each round we allow an analyst to ask a number of queries proportional to the number of queries they could ask in the independent setting. \par 
Unfortunately, the round-robin scheduler depends on each analyst always having a query ready when it is their turn to answer. In order to preserve the sharing incentive if an analyst does not have a query ready the Scheduler must stall and wait for their query. In practical settings, this could lead to long times waiting while the mechanism is stalled. 
\subsection{Randomized Scheduler}

The randomized scheduler, like the round-robin scheduler, selects an analyst at each time step to answer a query. Unlike the round-robin scheduler which answers queries in a deterministic order, the randomized scheduler randomly selects an analyst at each time step. The randomized scheduler is less restrictive than the round-robin scheduler while still ensuring that an efficient enough mechanism can satisfy the sharing incentive without any further changes. This can be demonstrated through a reduction to the well studied coupon collectors problem \cite{DoubleDixie,shank_yang_2013}.

In the coupon collectors problem, there are $k$ unique types of coupons in an urn. At each time step one coupon is chosen at random (with replacement). The goal is to find the value of $T_k(\vec{m}, \vec{p})$, that is the expected number of time steps required to acquire $\vec{m_i}$ of each coupon when there are $\vec{p_i}$ of each coupon in the urn.  

We can ask a similar question of the randomized scheduler. At each time step, given that an analyst is chosen at random to answer their query, in expectation how many queries must be answered so that each analyst answers at least as many queries as in the independent case. This reduces directly to the coupon collectors problem if you set $\vec{m}$ to be the number of queries that each analyst receives in the independent case $T_k(\vec{m},\vec{p})$ becomes the expected number of queries required for each analyst to at least answer as many queries as in the independent case. As such we can prove \cref{Thrm:RandomUniform}.\par

\begin{theorem}[Uniform Randomized Scheduler]
\label{Thrm:RandomUniform}
Let each of the $k$ analysts have an equal share of the privacy budget. Assume that at each time step an analyst is selected uniformly at random to answer a query. Given a mechanism $\gM$ which can answer at most $c$ queries with expected error under threshold $\alpha$ with privacy budget $\epsilon$ and can answer at least the first $k( \log(k) +  (c - 1)\log(\log(k)) + o(1)) $ queries under threshold $\alpha$ with privacy budget $k \cdot \epsilon$ satisfies the sharing incentive.
\end{theorem}

\begin{proof}
If a random analyst is chosen at each round to answer a query by \cref{Coupons:Uniform} the expected number of queries required to satisfy the sharing incentive for all analysts is $T_k(c,1) =  k( \log(k) +  (m - 1)\log(\log(k)) + o(1)) $. 
\end{proof}
 We can use \cref{Coupons:General} to extend this statement to hold for any distribution of privacy budgets and non-uniform distribution.
 
\begin{theorem} [Non-Uniform Randomized Scheduler] \label{Thrm:RandomGeneral}
\new{Let $\vec{p}$ be the vector with values proportional to the probabilities that each analyst is chosen during any time step and $\vec{m}$ be the vector that states the number of queries each analyst can answer independently. Given a mechanism $\gM$ which can answer at most $\vec{m_i}$ queries with expected error under threshold $\alpha$ with privacy budget $s_i\epsilon$ and can answer at least the first $\frac{p_{max}}{p_{min}} k( \log(k) +  (m_{max} - 1)\log(\log(k)) + o(1)) $ queries under threshold $\alpha$ with privacy budget $ \epsilon$ satisfies the sharing incentive, where $p_{max}$ and $p_{min}$ are the maximum and minimum values of $\vec{x}$ respectively and $m_{max}$ is the maximum of $\vec{m}$.}
\end{theorem}

 \begin{proof}
 First we note that  \cref{Coupons:Upper} upper bounds $T_k(\vec{m},\vec{p}) \leq \frac{p_{max}}{p_{min}}T_k(m_{max},1) $. We then once again apply \cref{Coupons:Uniform} directly to get $\frac{p_{max}}{p_{min}} k( \log(k) +  (m_{max} - 1)\log(\log(n)) + o(1)) $ thus proving \cref{Thrm:RandomGeneral}.
 \end{proof}

We note that while \cref{Thrm:RandomGeneral} implies that any mechanism efficient enough to answer that many queries will always satisfy the sharing incentive it is not always the case that such a mechanism exists. For instance, if  $T_k(\vec{m},\vec{x})$ is large enough
 then  the lower bounds of  \cite{dinur_nissim_2003,hardt2009geometry} ensure that no mechanism can answer all the queries under threshold $\alpha$.

\begin{algorithm}[h]
\label{Algo:CacheAndReconstruct}
\SetAlgoLined
\SetKwInOut{Input}{input}\SetKwInOut{Output}{output}
\Input{Sequence of queries with associated analyst $\gQ \leftarrow \{(q_1,a_i) \dots (q_{|\gQ|},a_i) \}$,\\
Number of analysts $k$ \\
Differentially private mechanism $\gM$,\\
Set of parameters $\gP$, \\} 
\Output{Sequence of Query Answers}
\vspace{3 mm}
Initialize mechanism $\gM$ with parameters $\gP$\\
Create a buffer $\vec{B}$ of queries for each analyst \\ \vspace{3 mm}
\nonl \textbf{At each time step $i$}\\
\If{$i \leq |\gQ|$}{
Add $\gQ_i$ to the buffer of the associated analyst}
Sample $j \in [1,k]$ uniformly at random \\
\
\Return The first query in analyst $j$s buffer with $\gM$

 \caption{Randomized Scheduler}
\end{algorithm}

\subsection{Utility and Trade-offs}

Unlike seeded cache and reconstruct the query schedulers enforce the sharing incentive while circumventing the upper bound of \cref{Thrm:SharingIncentiveUpper}. The query scheduler inherits the efficiency guarantee of the online mechanism that it uses. This ensures that the query scheduler can answer just as many queries as a traditional online mechanism that doesn't satisfy any of the desiderata and can answer significantly more queries than independent mechanisms.\par 
In the standard online model mechanisms that satisfy the sharing incentive incur a penalty to the number of queries they can answer. The query scheduler instead incurs a penalty to how fast it can answer those queries.
Since the query scheduler enforces which analysts can ask queries at any given time it cannot progress if the chosen analyst has no queries to be answered. This causes the query scheduler to stall and wait for an analyst to ask more queries (or indicate that they are done asking queries), preventing any other analyst from asking their queries. \new{ We measure the impact of stalling in practice by measuring each mechanism's time to completion in \cref{experiments}}. \par 
The round-robin scheduler ensures that any efficient mechanism satisfies the sharing incentive however severely restricts the ordering of analysts. By ensuring that all analysts have asked the same amount of queries the round-robin scheduler is guaranteed to stall if any analyst has fewer queries prepared than any of the others. The randomized scheduler requires an underlying mechanism that is more efficient than the round-robin scheduler. In exchange, the randomized scheduler stalls less often as the order of analysts is sampled from a distribution which can be chosen to match some prior knowledge of analysts' behavior. We show in \cref{experiments} that this is particularly desirable when some analysts ask significantly fewer queries than others. In that case, even when the analyst chosen is sampled from the uniform distribution it incurs less stalling time than the round-robin scheduler.

\section{Experiments}
\label{experiments}

\begin{figure*}[t] 
\begin{subfigure}[b]{0.28\textwidth}
\includegraphics[width=\linewidth]{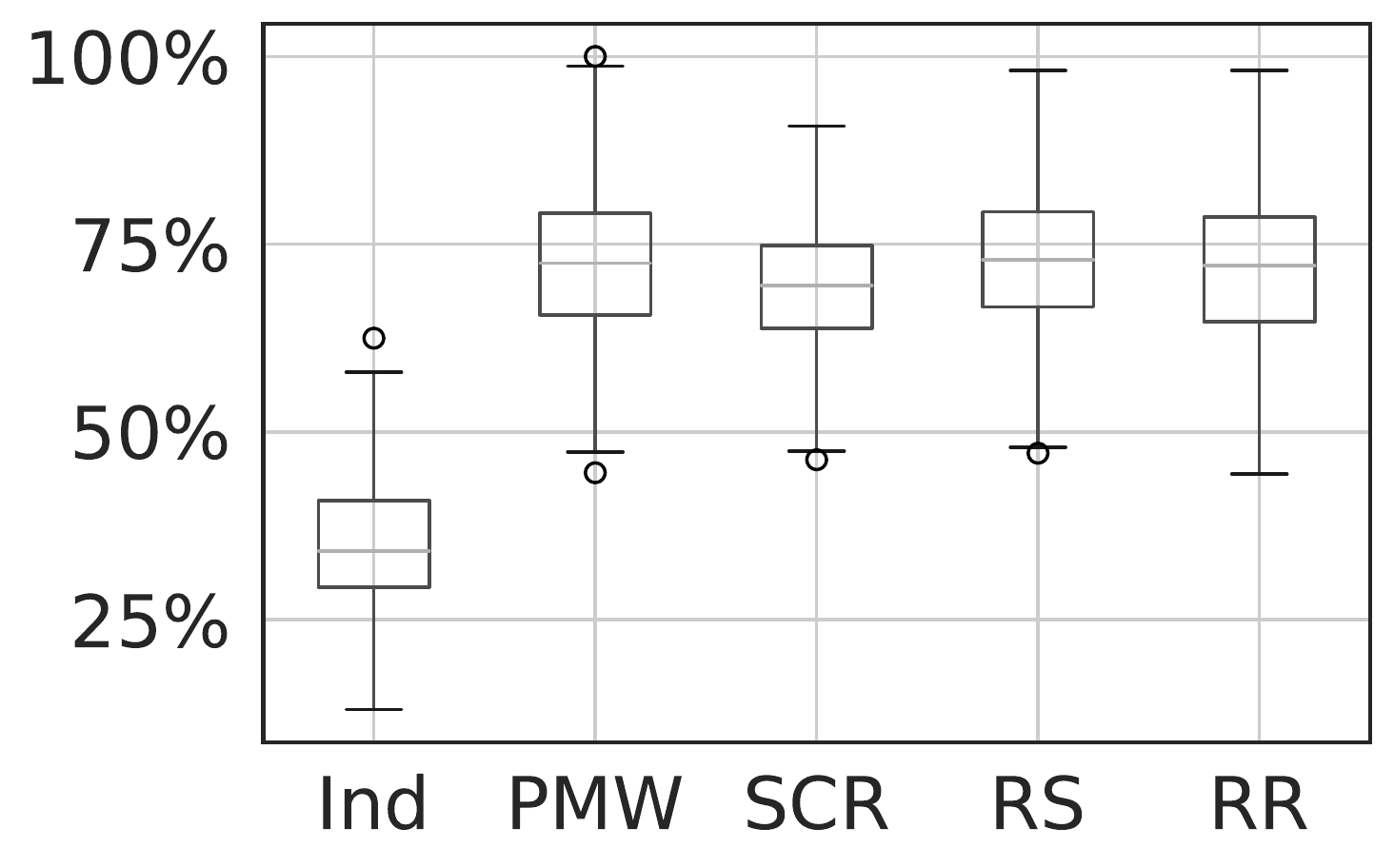}
\caption{Accuracy for p = 0.01} \label{fig:0.01acc}
\end{subfigure}\hspace*{\fill}
\begin{subfigure}[b]{0.28\textwidth}
\includegraphics[width=\linewidth]{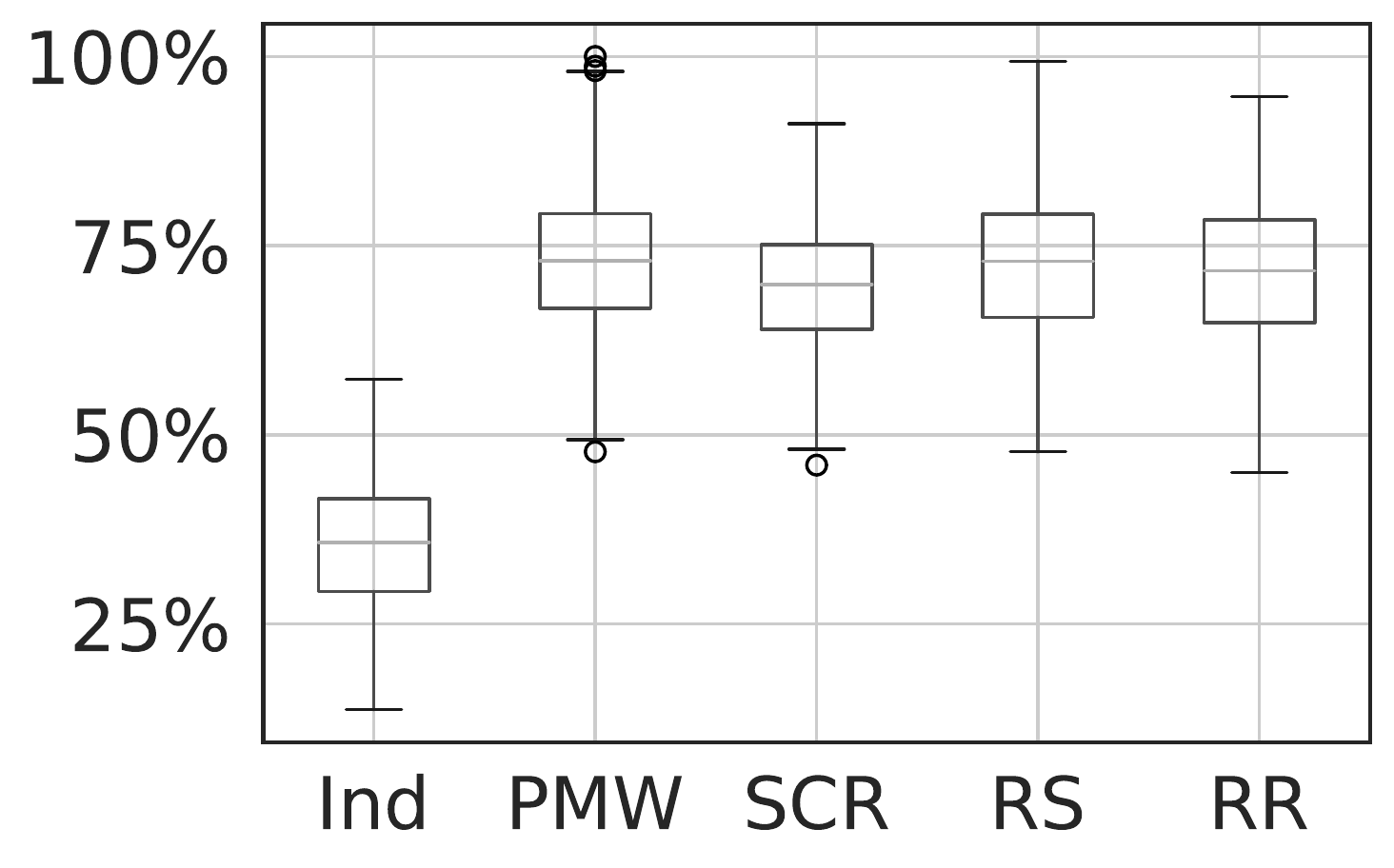}
\caption{Accuracy for p = 0.1} \label{fig:0.1acc}
\end{subfigure}\hspace*{\fill}
\begin{subfigure}[b]{0.28\textwidth}
\includegraphics[width=\linewidth]{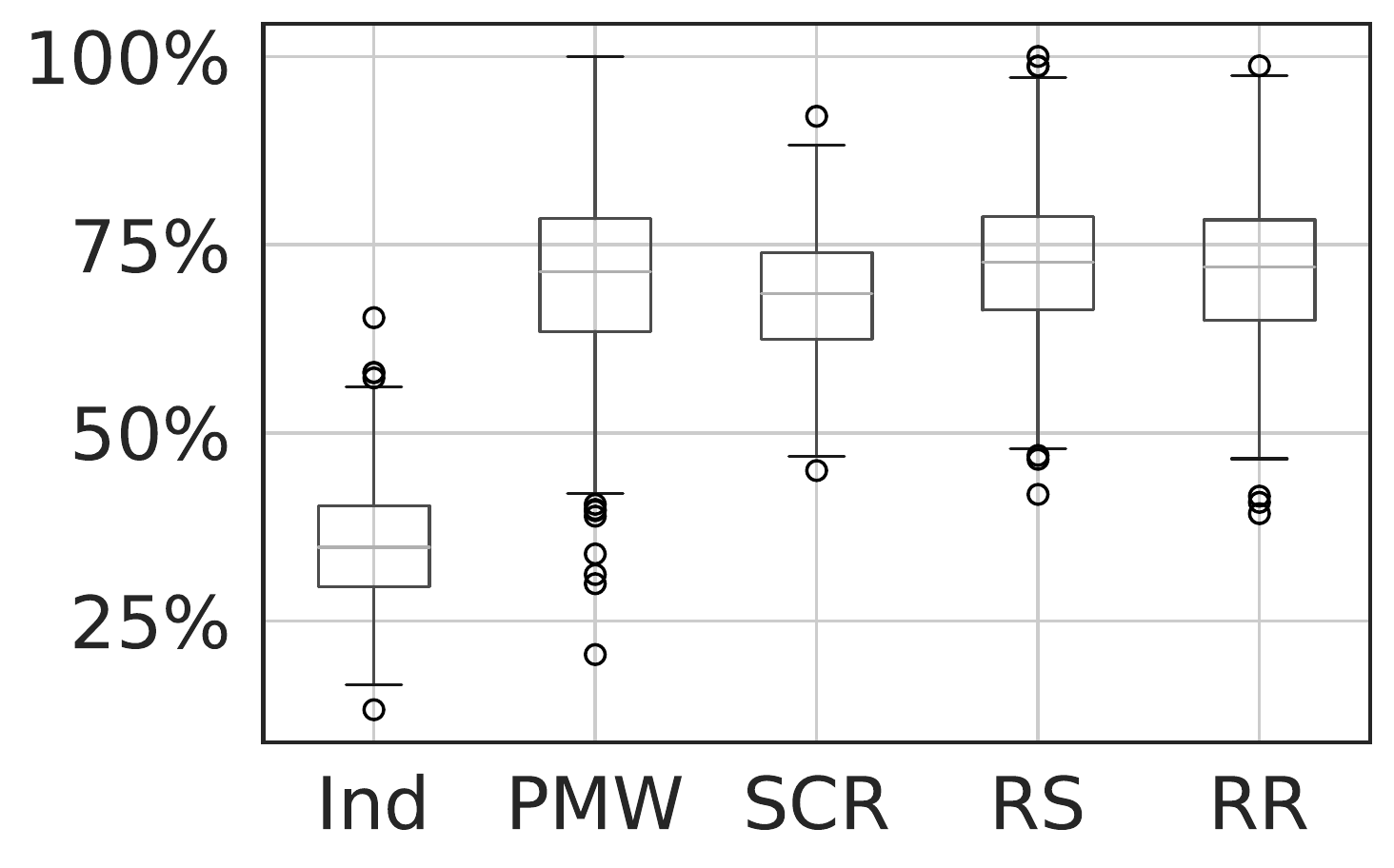}
\caption{Accuracy for p = 0.9} \label{fig:0.9acc}
\end{subfigure}

\medskip
\begin{subfigure}[b]{0.28\textwidth}
\includegraphics[width=0.92\linewidth, right]{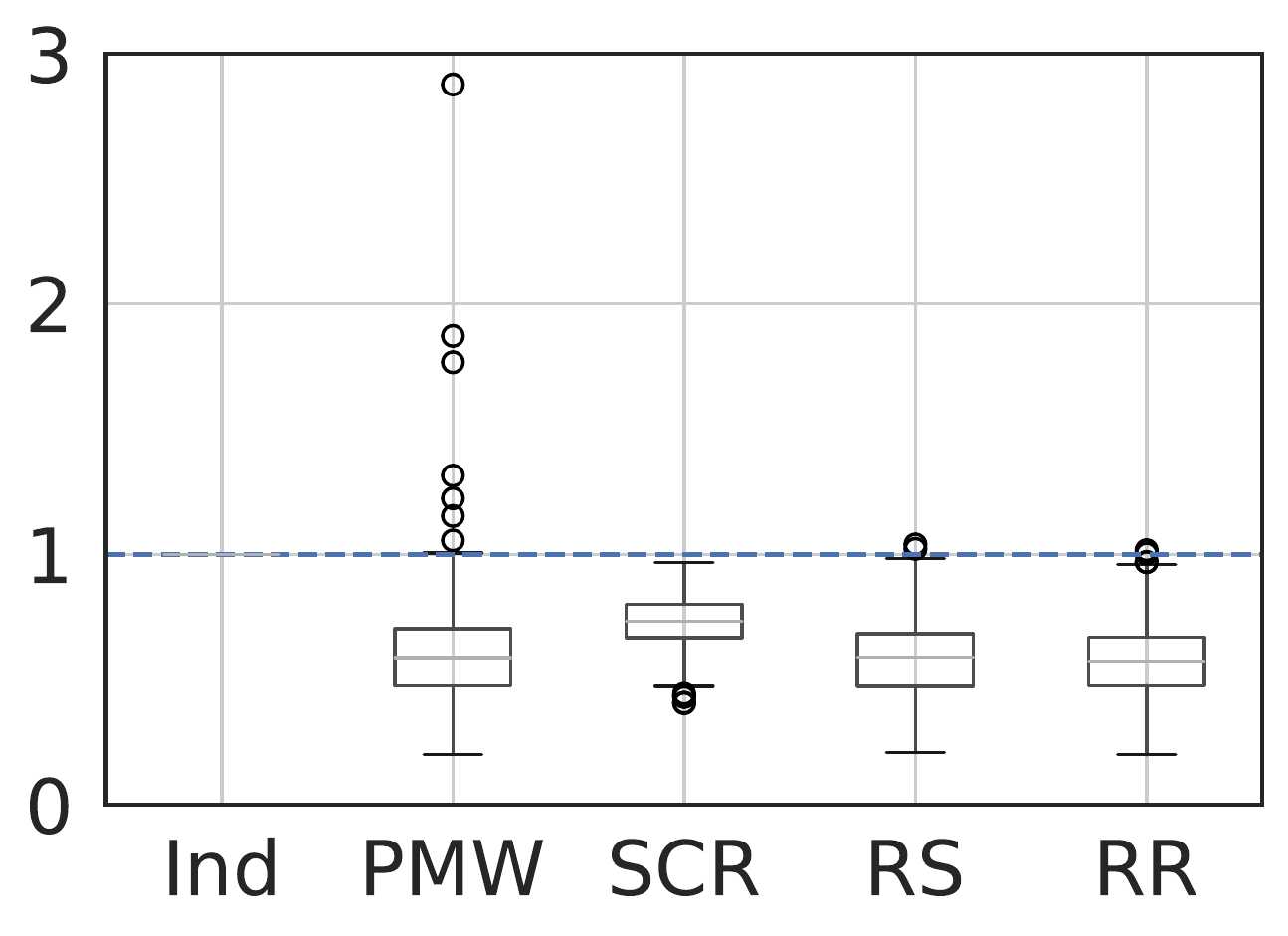}
\caption{Max Ratio for p = 0.01} \label{fig:0.01ratio}
\end{subfigure}\hspace*{\fill}
\begin{subfigure}[b]{0.28\textwidth}
\includegraphics[width=0.92\linewidth, right]{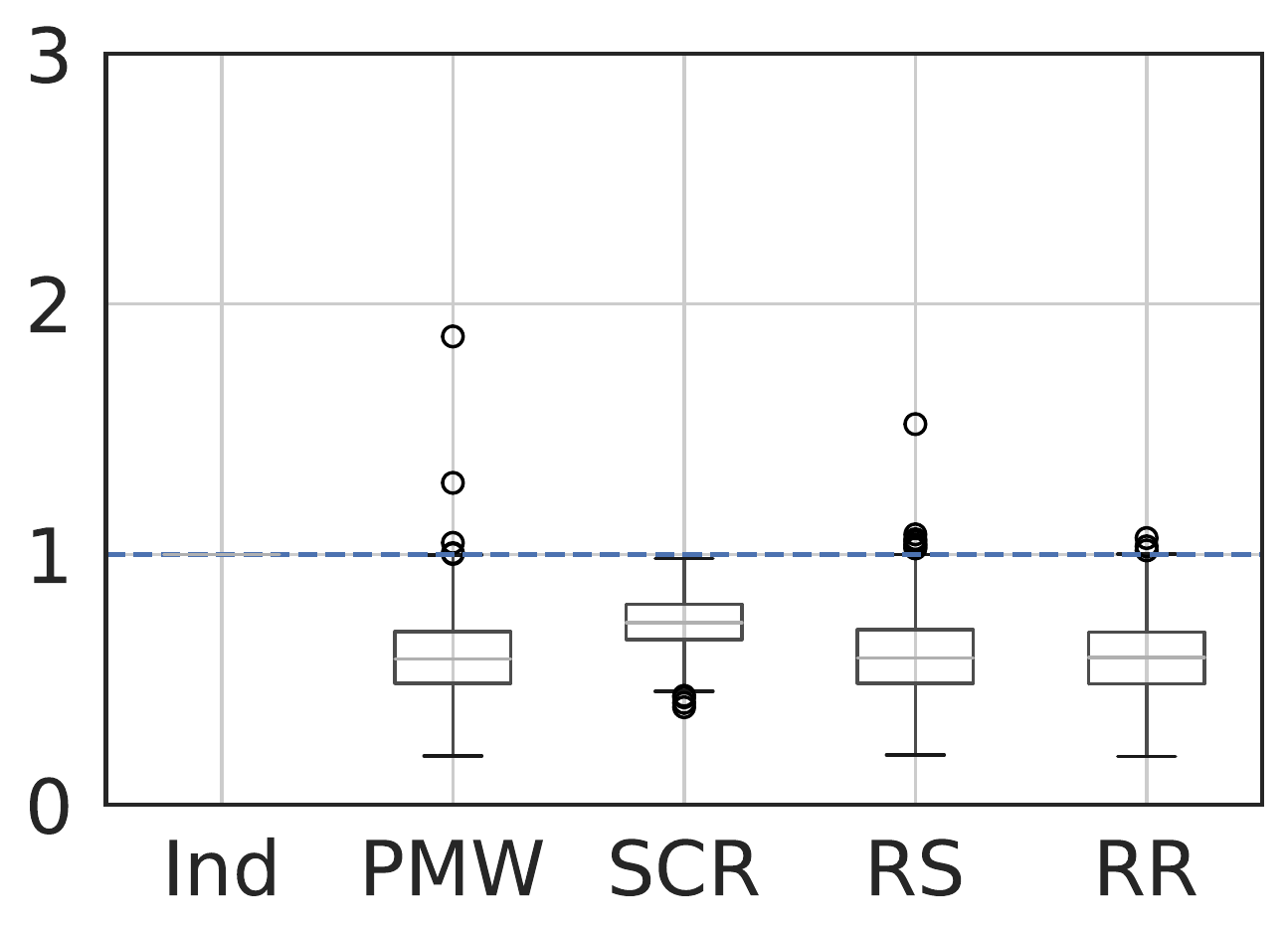}
\caption{Max Ratio for p = 0.1} \label{fig:0.1ratio}
\end{subfigure}\hspace*{\fill}
\begin{subfigure}[b]{0.28\textwidth}
\includegraphics[width=0.92\linewidth, right]{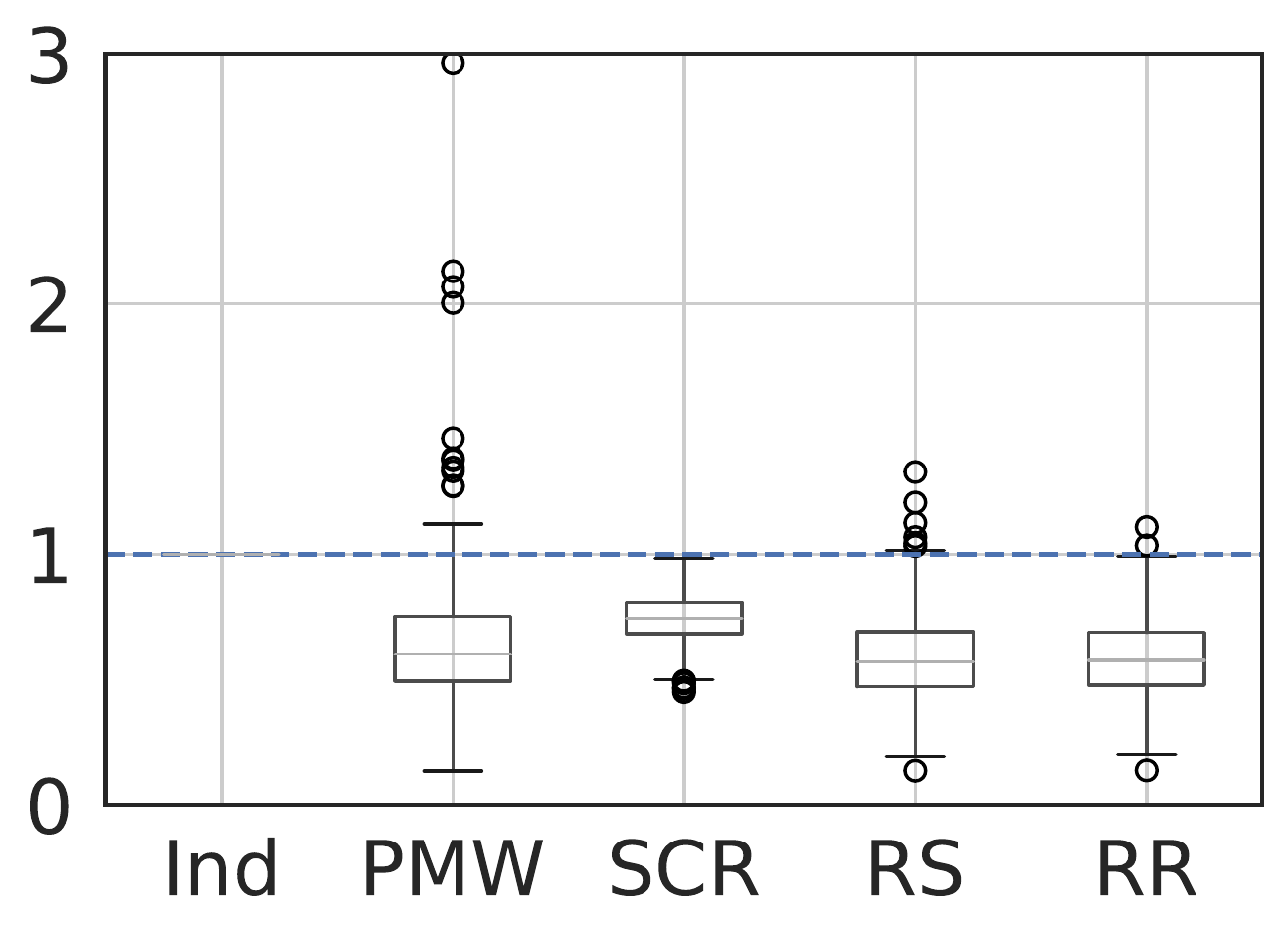}
\caption{Max Ratio for p = 0.9} \label{fig:0.9ratio}
\end{subfigure}

\medskip
\begin{subfigure}[b]{0.28\textwidth}
\includegraphics[width=0.92\linewidth, right]{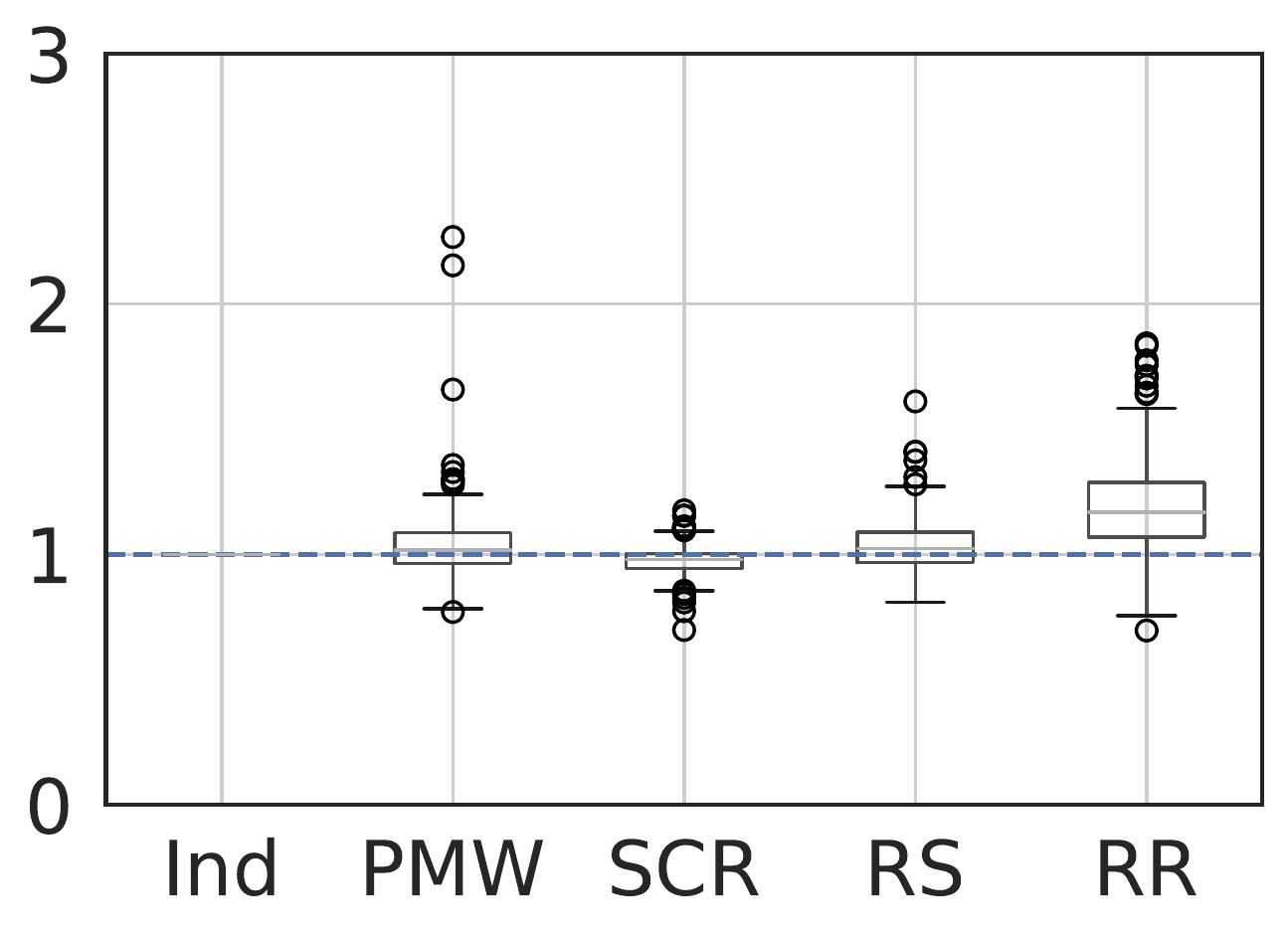}
\caption{ Interference for p = 0.01} \label{fig:0.01inter}
\end{subfigure}\hspace*{\fill}
\begin{subfigure}[b]{0.28\textwidth}
\includegraphics[width=0.92\linewidth, right]{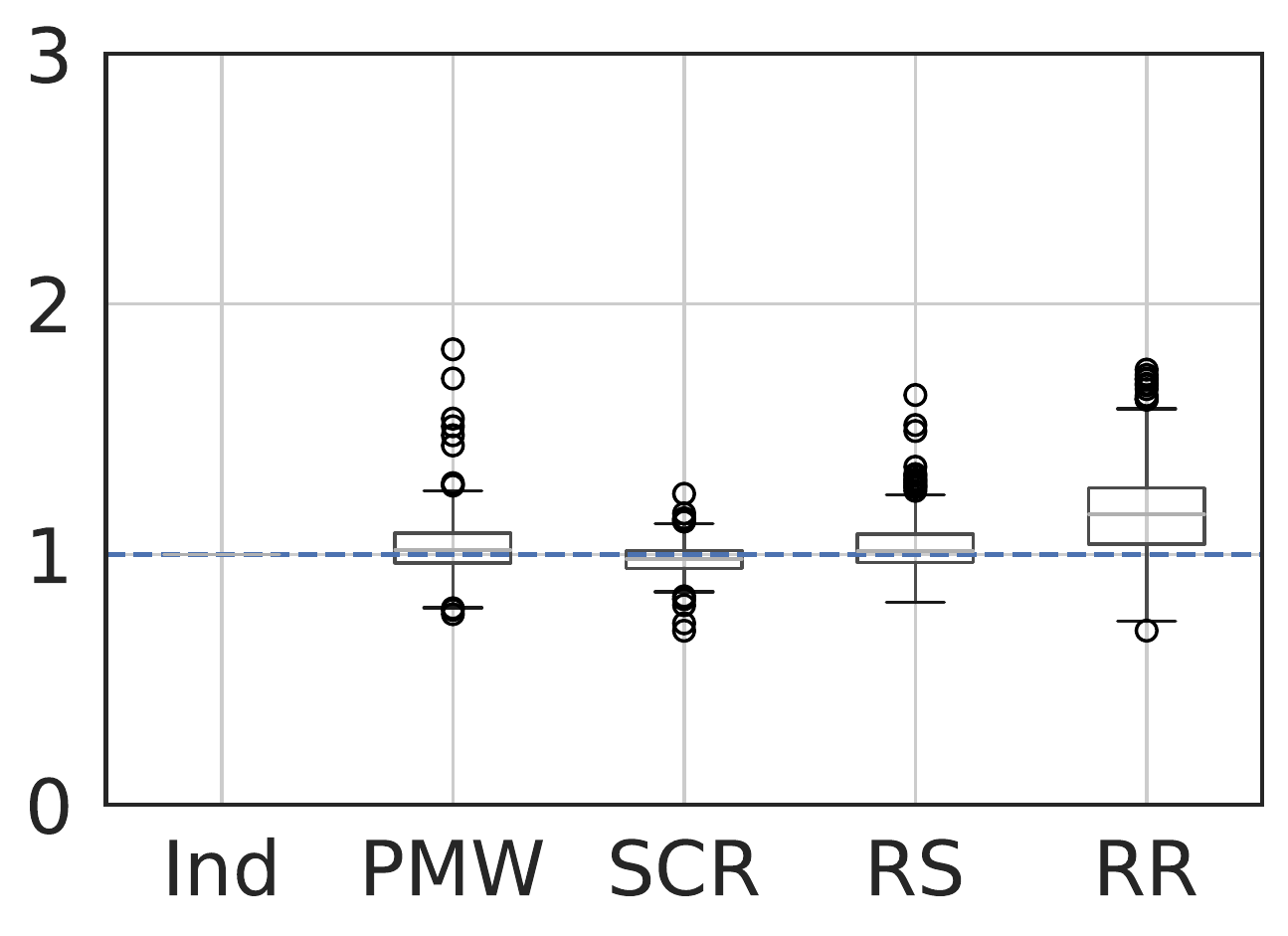}
\caption{ Interference for p = 0.1} \label{fig:0.1inter}
\end{subfigure}\hspace*{\fill}
\begin{subfigure}[b]{0.28\textwidth}
\includegraphics[width=0.92\linewidth, right]{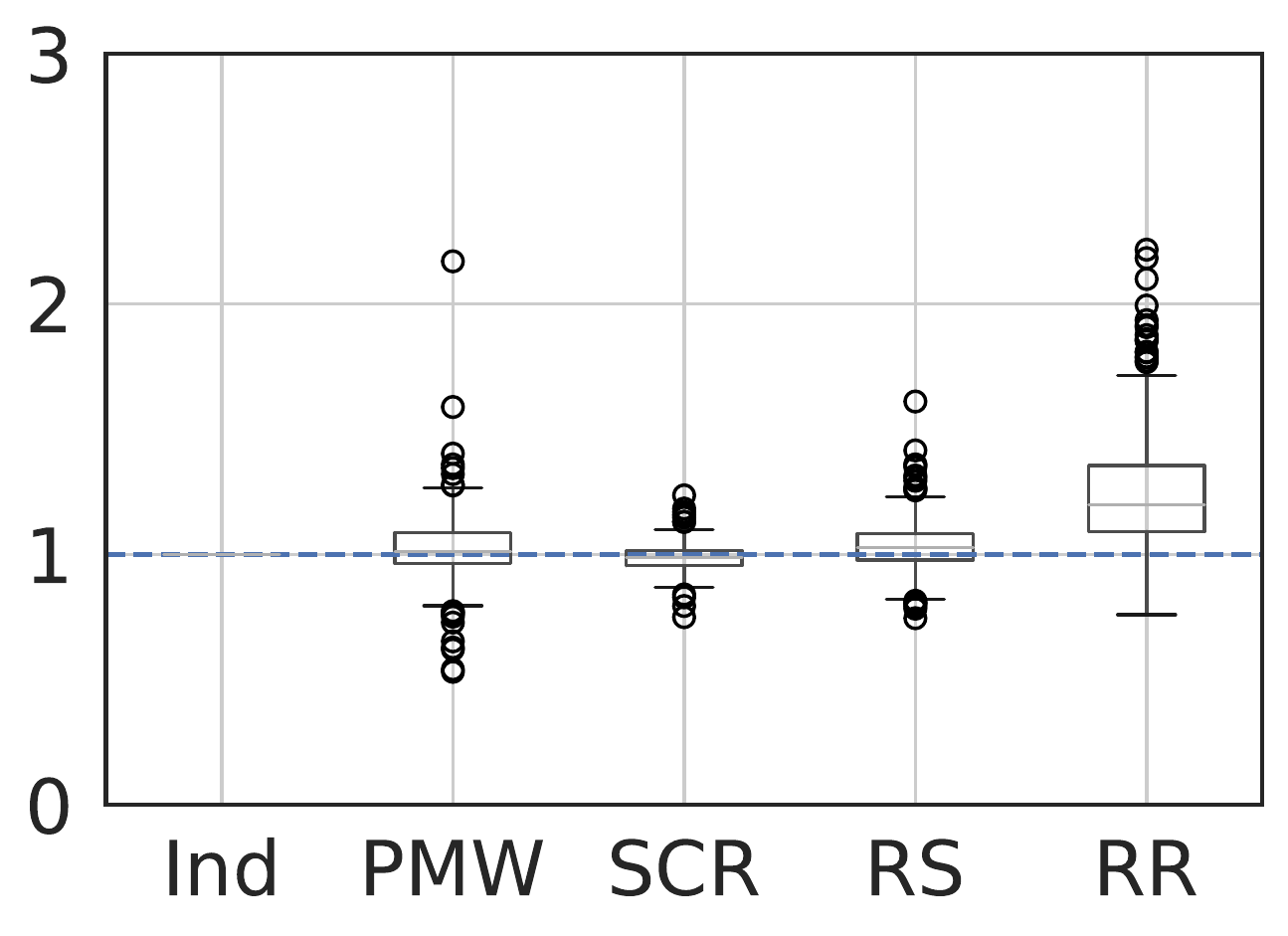}
\caption{ Interference for p = 0.9} \label{fig:0.9inter}
\end{subfigure}
\medskip
\begin{subfigure}[b]{0.28\textwidth}
\includegraphics[width=\linewidth]{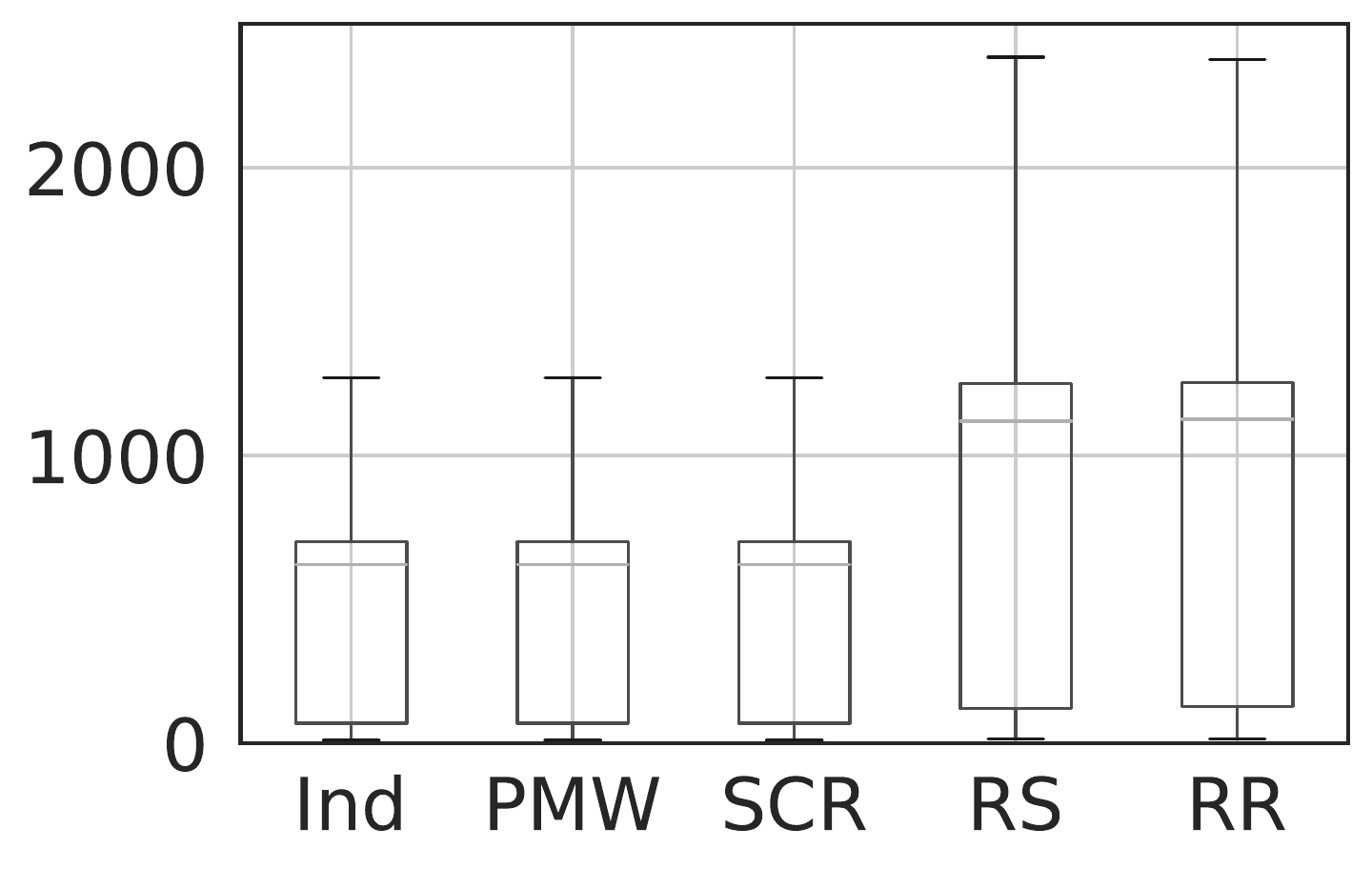}
\caption{Time Steps for p = 0.01} \label{fig:0.01time}
\end{subfigure}\hspace*{\fill}
\begin{subfigure}[b]{0.28\textwidth}
\includegraphics[width=\linewidth]{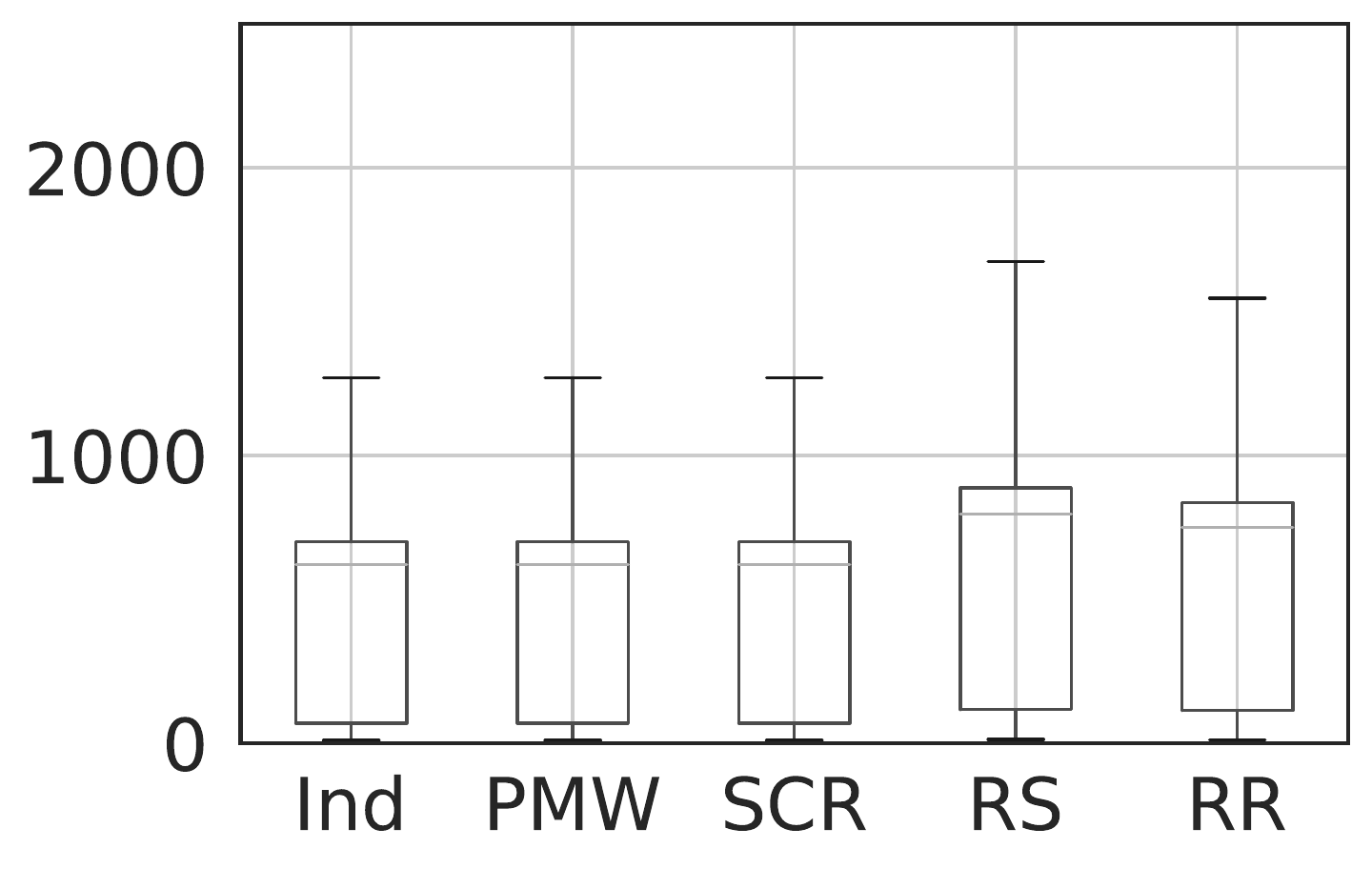}
\caption{Time Steps for p = 0.1} \label{fig:0.1time}
\end{subfigure}\hspace*{\fill}
\begin{subfigure}[b]{0.28\textwidth}
\includegraphics[width=\linewidth]{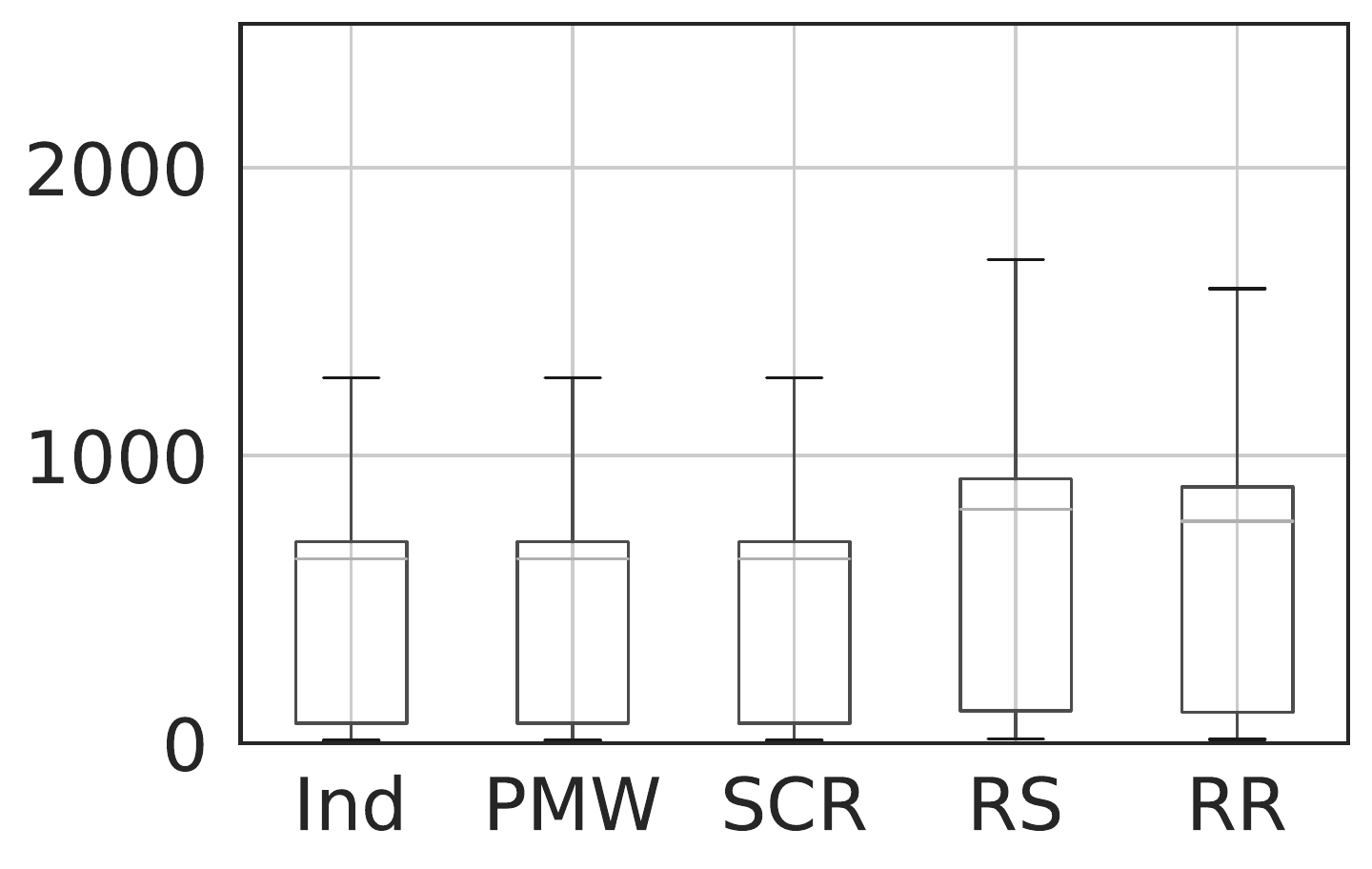}
\caption{Time Steps for p = 0.9} \label{fig:0.9time}
\end{subfigure}

\caption{ Accuracy (First row), Max Ratio Error (Second row), Empirical Interference (Third row), and Time to Completion (Last row)  for the following five algorithms: Independent PMW(Ind), Joint PMW (PMW), Seeded Cache and Reconstruct (SCR), Round Robin PMW (RR), and Randomized PMW (RS).
p represents the probability that the first analyst (among 10) has her queries answered at any time step. 
} \label{fig:1}
\end{figure*}
\new{
We design experiments to both test if the mechanisms proposed satisfy the desiderata as well as how well they perform in practice. The randomized process used to generate sequences is designed to emulate possible sequences across multiple analysts that one might see in a practical setting. 
}

\subsection{\new{Experimental Setup}}
The following experiments are largely similar to the experiments in \cref{sec:motivation} which have been extended to the case of more than 2 analysts. We consider 10 analysts each with equal shares of the privacy budget $\epsilon/10$. \new{This allows for more interactions and more complex interactions between analysts than in the 2 analyst case.} \new{We evaluated several different privacy budgets $\epsilon \in  [0.1, 1 , 10]$ but found that the results remained largely the same across privacy budgets. As such all the results shown below are using a moderate privacy budget $\epsilon = 1$.}\par   
A query sequence is generated by first assigning each analyst with a workload from a list of either one of the census race workloads \cite{HDMM}, the identity workload, prefix sum, or H2 workload.
Each of these workloads can either be asked on the entire database or a subset of the database \new{defined by a predicate}. \new{This ensures that there is a mix of overlapping and disjoint queries.} These workloads are then merged by randomly selecting an analyst to answer a query at each time step.
Like in \cref{sec:motivation} there is a parameter $p$ which denotes the probability that the first analyst is selected to answer a query. The first analyst is chosen with probability $p$ \new{each of the other analysts are chosen with probability $(1-p)/9$}. In this case, since there are 10 analysts $p= 0.1$ corresponds to the case where an analyst is chosen uniformly at random each time and values of $p$ above that signify that the first analysts asks their queries with higher probability.
We include $p$ values of $0.01$ , $0.1$,  and $0.9$.
This includes a case when one analyst is underrepresented, a uniform distribution, and a case where one analyst is vastly over represented.
\subsection{Mechanisms}

We use \textbf{Independent PMW} (ind) as our baseline mechanism which we will compare our other mechanisms to. In this mechanism, each analyst is given their own independent instance of PMW and asks their queries exclusively on their respective instance. Since this mechanism does not allow any interaction between analysts it satisfies the sharing incentive, analyst monotonicity, and non-interference. \par 
We also compare our mechanisms against a mechanism that is designed to optimize for overall utility without any regard for the desiderata. For this purpose, we also evaluate \textbf{Private Multiplicative Weights} (PMW) as a general online differentially private mechanism, without regard to analyst identity.  
\par 
The first of our proposed mechanisms is \textbf{Seeded Cache and Reconstruct} (SCR), explained in detail in \cref{Sec:CacheAndReconstruct}. 
In addition, we test the efficacy of both the randomized scheduler and the round robin scheduler. For both schedulers, we create a single instance of PMW and pass that into the scheduler as a parameter. As a result, we evaluate \textbf{Round Robin PMW} (RR) and \textbf{Randomized Scheduler PMW} (RS) as instances of the round robin scheduler and randomized scheduler respectively. 

\subsection{Empirical Measures}
\label{sec:measures}
We measure overall utility as the number of queries that can be answered with error under a threshold $\alpha$. For each analyst, their individual utility is measured as the number of queries belonging to that analyst that are answered with error under threshold $\alpha$.\par 
In addition to utility, we measure the Maximum Ratio Error and Empirical Interference as in \cite{Waterfilling}. \new{These are measures of how severe any violations of the sharing incentive and non-interference respectively.} 
The ratio error of a mechanism $\gM$ of a given analyst $i$ is the utility of $\gM$ in the independent case divided by the utility of the $\gM$ in the joint case. This value measures the sharing incentive and to what extent it is violated. Values greater than 1 signify a violation of the sharing incentive and larger values signify larger violations. Here we present the \textbf{Maximum Ratio Error} which is the maximum of all ratio errors taken across all analysts. 
\begin{equation*}
    \max_i \left(  \frac{U_i(\gM,\gQ^i, s_i\epsilon)}{U_i(\gM,\gQ, \epsilon)} \right)
\end{equation*} 

The Empirical Interference is a measure of the extent to which a mechanism violates non-interference. For any analyst $i$ the interference with respect to another analyst $j$ is measured as the ratio of utility of analyst $i$ under mechanism $\gM$ excluding $j$ and the utility of analyst $i$ under the same mechanism with all analysts present. Like before if this ratio is larger than 1 then analyst $i$ experiences more utility when $j$ is excluded than when they are included in the mechanism, indicating a violation of non-interference. We define the Empirical Interference as the maximum interference across any pair of analysts as follows. 
\begin{equation*}
\max_{i,j, i \neq j} \frac{U_i(\gM,\gQ \setminus \gQ^j, (1-s_j)\epsilon)}{U_i(\gM,\gQ, \epsilon)}
\end{equation*} 

The Query Schedulers incur an additional cost in that they can stall when queries are not available. In order to measure the impact of the schedulers' stalling, we measure the \new{Time To Completion}, the number of time steps necessary for a mechanism to answer all the queries. For the non-scheduled mechanisms this will simply measure the size of the query sequence but for the scheduled mechanisms this will capture the stalling time in addition to the time answering the queries. 

\subsection{Results} \label{sec:experiment_results}
We see in \cref{fig:1} the results for $\epsilon =1 $ and $\alpha = 0.01$.
\subsubsection{Utility}
In terms of utility, it is clear that independently answering queries results in a severe decrease in utility. Independent PMW can answer less than half of the total queries under the threshold while all of the other mechanisms perform significantly better under all values of $p$. \par 
While SCR is analyst monotonic in all cases it comes at a slight cost. SRC consistently answers slightly fewer queries than the optimal PMW or any of the schedulers but significantly better than independent mechanisms. \new{Since the schedulers only re-order the query sequence they perform nearly identically to PMW. They even outperform PMW in pathological cases where one analyst's queries are over-represented at the beginning of the sequence when $p= 0.9$.}
\subsubsection{Max Ratio Error}
PMW regularly violates the sharing incentive for all values of $p$. \new{This is particularly severe in cases where one analyst is either over or under represented. In those cases, some analysts can see as many as 3 times more queries answered in the independent case as opposed to the joint case.}
SCR has no violations of the sharing incentive in any case whereas the schedulers only observe violations in outlier cases due to the inherent randomness of those mechanisms.
\subsubsection{Empirical Interference}
In most cases, PMW sees a violation of non-interference, with only rare cases having no violations. Both schedulers also violate non-interference in most cases and the violation grows as $p$ increases. Of the two schedulers, the randomized scheduler observes significantly less severe violations of the sharing incentive. For example in \cref{fig:0.9inter} in the worst case, an analyst can only answer $1.5 \times$ the queries without one analyst whereas the round robin scheduler sees a $2.2\times$ difference in the worst case. SCR sees only a few outlier violations of the sharing incentive which can be attributed to randomness. \new{Unlike the schedulers, which do not provably satisfy the sharing incentive, SCR sees only minor violations and still satisfies the sharing incentive in expectation.}
\subsubsection{Time to Completion}
Time to completion remains the same for all the non-schedulers as there is no possibility for stalling. \new{In these cases, the time to completion is simply the number of queries in the sequence}. In all cases, both schedulers incur some penalty in time to completion. This is the most extreme in the case of $p=0.01$ when one analyst is severely underrepresented. This can lead to frequent stalling where the mechanism is waiting on the underrepresented analyst. \new{In the worst case, this leads to the schedulers taking up to twice as many iterations to completely answer the entire sequence.}

\subsection{\new{Discussion}}

\new{We demonstrated that both seeded cache and reconstruct and the query schedulers are efficient solutions that both satisfy the sharing incentive. While they are both viable mechanisms each serves its own purpose and the choice of which to use is left to the data curator. Seeded cache and reconstruct has the benefit of being provably analyst monotonic and as such can be used in high stakes cases where satisfying the criteria is crucial. The query schedulers, however, have the benefit that they can be built on top of existing state of the art mechanisms and are not subject to the fundamental query limit. These can be used in lower stakes cases where the additional utility from the current state of the art outweighs the need for the guarantees of non-interference.} 

\section{Conclusion}
\label{Conclusions}

We demonstrate through \cref{Thrm:SharingIncentiveUpper} that  online mechanisms can either answer a large number of queries (like PMW) or satisfy the sharing incentive when the ordering is possibly adversarial. This result relies heavily on the ability for queries to arrive in an adversarial order where one analyst monopolizes those first set of queries. We first propose Seeded Cache and Reconstruct, a mechanism which is analyst monotonic in all cases but is subject to the limit of \cref{Thrm:SharingIncentiveUpper}. We then propose the alternative Query Scheduler which allows existing state-of-the-art online query answering mechanisms to satisfy the sharing incentive without compromising performance.  
\begin{acks}
This work was supported by the NSF award NSF SATC-2016393.
\end{acks}
\bibliographystyle{ACM-Reference-Format}
\bibliography{sample}

\end{document}